\documentclass[journal=jpclcd,manuscript=letter, layout=twocolumn]{achemso}
\setkeys{acs}{super=false}
\setcitestyle{numbers}

\usepackage[version=3]{mhchem} % Formula subscripts using \ce{}
\usepackage[dvipsnames]{xcolor}
\usepackage{xspace}
\usepackage{multirow}
\usepackage{array}

\usepackage[colorlinks=true,
    linkcolor=blue,
    filecolor=blue,
    citecolor=blue,      
    urlcolor=blue,]{hyperref}

\usepackage{prettyref}

\newcommand{\pref}[1]{\prettyref{#1}}
\newrefformat{fig}{Fig.~\ref{#1}}
\newrefformat{tab}{Table~\ref{#1}}
\newrefformat{sec}{Sec.~\ref{#1}}
\newrefformat{ssec}{Sec.~\ref{#1}}
\newrefformat{app}{App.~\ref{#1}}
\newrefformat{eqn}{Eq.~(\ref{#1})}

\newcolumntype{R}[1]{>{\raggedleft\let\newline\\\arraybackslash\hspace{0pt}}m{#1}}

\newcommand{\nbcl}{Nb$_3$Cl$_8$\xspace}
\newcommand{\nbbr}{Nb$_3$Br$_8$\xspace}
\newcommand{\nbi}{Nb$_3$I$_8$\xspace}
\newcommand{\nbx}{Nb$_3$$X_8$\xspace}
\newcommand{\nbclbr}{Nb$_3$Cl$_4$Br$_4$\xspace}
\newcommand{\nbclbrx}{Nb$_3$Br$_{4-x}$Cl$_{4+x}$\xspace}
\newcommand{\dftuv}{DFT$+U$$+V$\xspace}

\author{Alberto Carta}
\altaffiliation{These authors contributed equally to this work.}
\email{alberto.carta@psi.ch}
\affiliation{PSI Center for Scientific Computing, Theory, and Data, 5232 Villigen PSI, Switzerland}
\alsoaffiliation{Materials Theory, ETH Z\"urich, Wolfgang-Pauli-Strasse 27, 8093 Z\"urich, Switzerland}
\author{Peter Mlkvik}
\altaffiliation{These authors contributed equally to this work.}
\affiliation{Materials Theory, ETH Z\"urich, Wolfgang-Pauli-Strasse 27, 8093 Z\"urich, Switzerland}
\author{Fabian Grahlow}
\affiliation{Section for Solid State and Theoretical Inorganic Chemistry, Institute of Inorganic Chemistry, Eberhard-Karls-Universit\"at T\"ubingen, Auf der Morgenstelle 18, 72076 T\"ubingen, Germany}
\author{Markus Str\"obele}
\affiliation{Section for Solid State and Theoretical Inorganic Chemistry, Institute of Inorganic Chemistry, Eberhard-Karls-Universit\"at T\"ubingen, Auf der Morgenstelle 18, 72076 T\"ubingen, Germany}
\author{H.-J\"urgen Meyer}
\affiliation{Section for Solid State and Theoretical Inorganic Chemistry, Institute of Inorganic Chemistry, Eberhard-Karls-Universit\"at T\"ubingen, Auf der Morgenstelle 18, 72076 T\"ubingen, Germany}
\author{Carl P. Romao}
\affiliation{Department of Materials, Faculty of Nuclear Sciences and Physical Engineering, Czech Technical University in Prague, Trojanova 13, Prague 120 00, Czech Republic}
\author{Nicola A. Spaldin}
\affiliation{Materials Theory, ETH Z\"urich, Wolfgang-Pauli-Strasse 27, 8093 Z\"urich, Switzerland}
\author{Claude Ederer}
\affiliation{Materials Theory, ETH Z\"urich, Wolfgang-Pauli-Strasse 27, 8093 Z\"urich, Switzerland}

\title[title]{Hubbard dimer physics and the magnetostructural transition in the correlated cluster material Nb$_3$Cl$_8$}

\abbreviations{...}
\begin{document}

% making inline citations have square brackets
\setcitestyle{square}

%%%%%%%%%%%%%%%%%%%%%%%%%%%%%%%%%%%%%%%%%%%%%%%%%%%%%%%%%%%%%%%%%%%%%
%% The "tocentry" environment can be used to create an entry for the
%% graphical table of contents. It is given here as some journals
%% require that it is printed as part of the abstract page. It will
%% be automatically moved as appropriate.
%%%%%%%%%%%%%%%%%%%%%%%%%%%%%%%%%%%%%%%%%%%%%%%%%%%%%%%%%%%%%%%%%%%%%
\begin{tocentry}

% \begin{figure}
  \includegraphics[width=0.95\columnwidth]{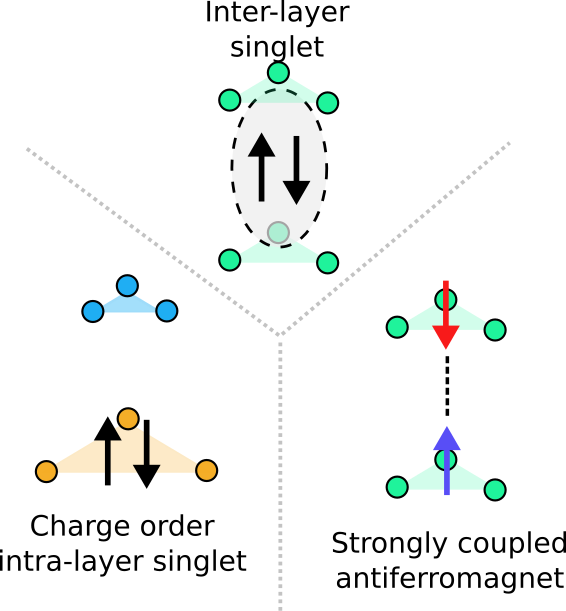}
  % \caption{This could be also the TOC figure}
  % \label{fig:example1}
% \end{figure}

% Some journals require a graphical entry for the Table of Contents. This should be laid out ``print ready'' so that the sizing of the text is correct. Inside the \texttt{tocentry} environment, the font used is Helvetica 8\,pt, as required by \emph{Journal of the American Chemical Society}. The surrounding frame is 9\,cm by 3.5\,cm, which is the maximum permitted for  \emph{Journal of the American Chemical Society} graphical table of content entries. The box will not resize if the content is too big: instead it will overflow the edge of the box. This box and the associated title will always be printed on a separate page at the end of the document.

\end{tocentry}

%%%%%%%%%%%%%%%%%%%%%%%%%%%%%%%%%%%%%%%%%%%%%%%%%%%%%%%%%%%%%%%%%%%%%
%% The abstract environment will automatically gobble the contents
%% if an abstract is not used by the target journal.
%%%%%%%%%%%%%%%%%%%%%%%%%%%%%%%%%%%%%%%%%%%%%%%%%%%%%%%%%%%%%%%%%%%%%
\begin{abstract}

We present a combined computational and experimental study of \nbcl, a correlated layered material containing Nb trimers, through the lens of competing intra- and intercluster interactions. Different proposed explanations for its magnetostructural transition such as charge disproportionation, antiferromagnetic quenching, and interlayer singlet formation are investigated in light of the various reported low-temperature structures. Our findings rule out the previously proposed charge-disproportionation, suggest an intricate interplay between Mott physics and the formation of interlayer singlets, and also hint at a possible explanation of the observed intratrimer scissoring distortion. We suggest that the physics of \nbcl should be understood in the context of weakly coupled Hubbard dimers.
\end{abstract}

%%%%%%%%%%%%%%%%%%%%%%%%%%%%%%%%%%%%%%%%%%%%%%%%%%%%%%%%%%%%%%%%%%%%%
%% Start the main part of the manuscript here.
%%%%%%%%%%%%%%%%%%%%%%%%%%%%%%%%%%%%%%%%%%%%%%%%%%%%%%%%%%%%%%%%%%%%%

\section{Introduction}

Cluster Mott insulators are strongly correlated materials for which the traditional definition of sites on which electrons interact~\cite{Roy:2019} is extended to include clusters of atoms~\cite{Muller/Kockelmann/Johrendt:2006, Kimber_et_al:2012, Nag_et_al:2016, Streltsov/Cao/Khomskii:2017, Jayakumar/Hickey:2023, Grahlow_et_al:2024}. The interplay of charge and spin degrees of freedom, combined with the complexity introduced by the crystal field of the cluster, makes these materials an exciting playground to explore and engineer the physics of strongly correlated systems.

Among these, Kagome cluster magnets~\cite{Nikolaev/Solovyev/Streltsov:2021} have garnered considerable interest due to the interplay between magnetism and correlation effects. In this work, we focus on \nbcl, part of the triniobium octahalide \nbx~($X=$ Cl, Br, I) family. These compounds have recently been recognized as possible spin-liquid candidates~\cite{Hu_et_al:2023, Liu_et_al:2024} and as obstructed-atomic-insulator candidates~\cite{Xu_et_al:2024a}; they exhibit thickness-dependent conductivity~\cite{Yoon_et_al:2020}, display the anomalous valley Hall effect~\cite{Peng_et_al:2020, Feng/Yang:2023}, and have been showcased in Josephson diodes~\cite{Wu_et_al:2022} and infrared sensors~\cite{Oh_et_al:2020}.

\begin{figure*}[htb]
  \includegraphics[width=0.9\textwidth]{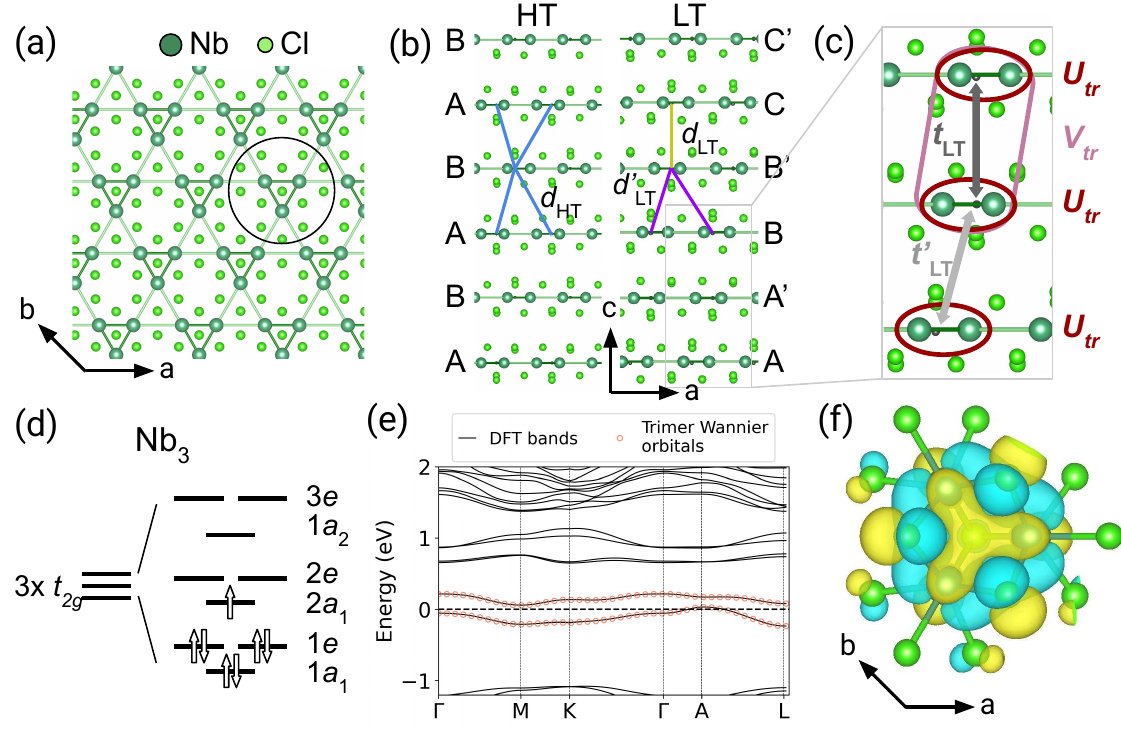}
  \caption{(a) Top view of the monolayer \nbcl with a trimer unit circled. (b) Side view of the bulk \nbcl in its HT and LT structures, showcasing the stacking and the change in distances between closest trimer centers across adjacent layers. (c) Closeup of the LT structure with the two relevant interlayer hoppings $t_{\text{LT}}$ and $t_{\text{LT}}^\prime$ indicated. $U_\text{tr}$ and $V_\text{tr}$ represent the intra- and inter-trimer interactions considered in our calculations. (d) Energy level diagram of the \nbcl trimer. (e) DFT Band structure of LT bulk \nbcl with the Wannier bands superimposed. (f) Top view of the corresponding Wannier function for one of the trimers within an LT unit cell.}
  \label{fig:intro}
\end{figure*}

\nbcl is a layered 2D material consisting of Nb$_3$Cl$_{13}$ cluster units [circled in \pref{fig:intro}(a)] arranged in a breathing Kagome lattice~\cite{Schaefer/Schnering:1964, Magonov_et_al:1993, Stroebele_et_al:2001} with short intratrimer bond distances of $d_{\text{Nb-Nb}} \cong 2.8$\,Å. The nominal electronic filling of each trimer is 7 electrons and due to the octahedral coordination of the Nb ions in the cluster crystal field, the energy levels are split, leaving a lone electron in the $2a_1$ orbital [as shown in \pref{fig:intro}(d)].
In both the monolayer and bulk forms, these $a_1$ orbitals form narrow bands close to the Fermi level which are well separated from the rest of the electronic bands as shown in \pref{fig:intro}(e). Wannier functions corresponding to these $a_1$ states can be constructed as triangular molecular orbitals of mixed Nb-halogen character which are extended over the whole trimer~\cite{Grytsiuk_et_al:2024} as seen in \pref{fig:intro}(f).

While the free-standing monolayer \nbcl has been characterized as a 2D Mott insulator both experimentally~\cite{Gao_et_al:2023, Shan_et_al:2023, Nakamura_et_al:2024} and theoretically~\cite{Zhang_et_al:2023, Gao_et_al:2023, Hu_et_al:2023, Grytsiuk_et_al:2024, Stepanov:2024}, the nature of the bulk material remains less explored. At high-temperature (HT), bulk \nbcl is found in the $\alpha$-phase with space group $P\bar{3}m1$, with two consecutive layers stacked in an AB-staggered pattern~\cite{Kennedy_et_al:1996, Stroebele_et_al:2001, Haraguchi_et_al:2017, Yoon_et_al:2020, Jiang_et_al:2022, Wu_et_al:2022}. The interlayer distance is $\sim 6.8$\,Å while the intertrimer distance (distance between centers of adjacent Nb$_3$ trimers in the stacking direction) is $d_{\text{HT}} \cong 7.8$\,Å~\cite{Haraguchi_et_al:2017} [\pref{fig:intro}(b), left]. 

As the temperature is lowered below $T_c \cong 100$\,K, the material remains insulating but undergoes a magnetostructural transition to the low-temperature (LT) $\beta$-phase which exhibits a significant change in the stacking pattern to an -AA'-BB'-CC'- arrangement~\cite{Haraguchi_et_al:2017, Pasco_et_al:2019, Sheckelton_et_al:2017}. This results in a ``dimerization'' of nearest-neighbor trimers along the stacking direction [\pref{fig:intro}(b), right]~\cite{Haraguchi_et_al:2017, Sheckelton_et_al:2017}. The newly formed pairs' intertrimer distance decreases significantly to $d_{\text{LT}} \cong 6.1$\,Å, and the calculated intertrimer hopping integral increases from 0.01\,eV to $t_{\text{LT}} \cong 0.15$\,eV~\cite{Grytsiuk_et_al:2024}. 

The space group of the LT $\beta$-\nbcl is still a matter of debate, with several different candidate structures proposed. Kim~\textit{et al.}~\cite{Kim_et_al:2023}, for instance, report $R\bar{3}m$ symmetry corresponding to the ideal version of this stacking, which has also been observed in other members of the extended Nb$_3$X$_8$ family: \nbclbr~\cite{Pasco_et_al:2019}, \nbbr ~\cite{Regmi:2023}, and \nbi~\cite{Kim_et_al:2023}. In the literature, lower symmetry structures such as $R3$~\cite{Haraguchi_et_al:2017, Haraguchi/Yoshimura:2024} and $C2/m$~\cite{Sheckelton_et_al:2017} have also been proposed. 

The different LT phases reported in the literature could plausibly be ascribed to differing experimental conditions used to synthesize and characterize \nbcl and related phases. The $R3$ and $C2/m$ phases have been observed using single crystal diffraction techniques~\cite{Haraguchi_et_al:2017, Sheckelton_et_al:2017}, which are generally highly sensitive to crystallographic symmetry, whereas $R\bar{3}m$ was assigned to the LT phase using a combination of experimental and theoretical vibrational spectroscopy ~\cite{Kim_et_al:2023}, which is a less reliable method of determining the space group. \nbclbr was also observed to adopt an $R\bar{3}m$ structure at low temperature using single crystal diffraction~\cite{Pasco_et_al:2019}.

In all cases, the structural transition is accompanied by a significant drop in the magnetic susceptibility~\cite{Haraguchi_et_al:2017, Sheckelton_et_al:2017, Pasco_et_al:2019, Haraguchi/Yoshimura:2024}, indicating a strong quenching of the local trimer moments. Several interpretations have been proposed to explain the observed drop in magnetic susceptibility. Haraguchi~\textit{et al.}~\cite{Haraguchi_et_al:2017, Haraguchi/Yoshimura:2024} report a ``breathing'' effect in the Nb$_3$ trimers, with alternating layers of shorter and longer Nb-Nb bond lengths in the $R3$ structure from single-crystal X-ray diffraction (XRD). The authors propose charge disproportionation, $2[\text{Nb}_3]^7 \rightarrow [\text{Nb}_3]^8 + [\text{Nb}_3]^6$, leading to a formation of a singlet on the larger Nb trimers and causing a drop in the magnetic susceptibility~\cite{Haraguchi_et_al:2017}.

When comparing \nbcl to other materials that exhibit singlet formation or charge disproportionation and subsequent structural distortions, several issues arise with the charge disproportionation model. Haraguchi~\textit{et al.}~\cite{Haraguchi_et_al:2017, Haraguchi/Yoshimura:2024} report bond lengths of $d_{\text{Nb1-Nb1}} = 2.801(3)$\,Å and $d_{\text{Nb2-Nb2}} = 2.821(2)$\,Å for the two inequivalent trimers, resulting in a notably small Nb-Nb bond-length difference of just $0.02$\,Å. We can compare these small distortions with other materials where metal-metal bonding is important. Vanadium dioxide (VO$_2$), for instance, is a material undergoing a similar paramagnetic-to-nonmagnetic transition associated with a formation of a V-V singlet state~\cite{Morin:1959, Goodenough:1971}. VO$_2$ features a magnetostructural transition (coupled to a metal-insulator transition) contracting the original V-V bond by almost 10\% or $\sim 0.3$\,Å~\cite{Eyert:2002a}, roughly 15 times that observed in \cite{Haraguchi_et_al:2017} for \nbcl. 

Moreover, using neutron diffraction, Sheckelton~\textit{et al.}~\cite{Sheckelton_et_al:2017} observe a different LT \nbcl structure which still exhibits the same magnetic-to-nonmagnetic transition, but has the $C2/m$ space group. This phase is characterized by an isosceles triangle geometry within the Nb$_3$ trimers leading to bond distances of $d_{\text{Nb1'-Nb2'}} = 2.86(2)$\,Å and $d_{\text{Nb2'-Nb2'}} = 2.72(1)$\,Å~\cite{Sheckelton_et_al:2017}. However, the authors ascribe the change of intratrimer distances to secondary effects and speculate that it is the stacking of the $\beta$ phase that drives a formation of singlets between different layers, causing the observed drop in susceptibility~\cite{Sheckelton_et_al:2017}. 

Computationally, due to the small bandwidth and thus the potential importance of local electron-electron interactions, \nbcl has largely been studied using a combination of density-functional theory (DFT) and dynamical mean-field theory (DMFT)~\cite{Gao_et_al:2023, Zhang_et_al:2023, Grytsiuk_et_al:2024, Aretz_et_al:2025} or similar beyond-DFT methods~\cite{Hu_et_al:2023, Xiong/Zhang/Zunger:2025a}. The bulk HT $\alpha$-\nbcl has been labeled as a cluster-Mott insulator with many similarities to the 2D precursor material~\cite{Gao_et_al:2023, Hu_et_al:2023, Zhang_et_al:2023, Grytsiuk_et_al:2024}, consistent with the isolated local moments of the experimentally observed HT Curie-Weiss susceptibility. For the LT $\beta$-\nbcl phase, most works agree on the important role of the increase in interlayer coupling [$t_{\text{LT}}$ versus $t^\prime_{\text{LT}}$ in \pref{fig:intro}(c)], even though the specific proposals for the relevant physics vary. 

By evaluating the role of an intratrimer Coulomb interaction correction $U_{tr}$, studies in bilayer LT $\beta$-\nbcl~\cite{Gao_et_al:2023, Zhang_et_al:2023} conclude that it is the competition between the interlayer hopping, $t_{\text{LT}}$, and intratrimer Coulomb interaction that determines whether the system is an interlayer singlet insulator or a Mott insulator, with strong interlayer antiferromagnetic fluctuations. This conclusion has been bolstered by the studies on bulk systems from Grytsiuk, Aretz~\textit{et al.}~\cite{Grytsiuk_et_al:2024, Aretz_et_al:2025}. The same authors also report a change from a Mott insulating hypothetical Nb$_3$F$_8$ to a weakly correlated \nbi, due to the decreasing onsite Coulomb interaction $U_{tr}$ and the increasing interlayer hopping. Although the authors perform calculations of the experimentally observed $R3$ and the $C2/m$ structures~\cite{Grytsiuk_et_al:2024}, the origin and stability of the $\beta$ \nbcl low symmetry phases with respect to the $R\bar3m$ phase remains unclear. 

In our work, we employ both \dftuv~\cite{LeiriaCampoJr/Cococcioni:2010} and DFT+DMFT \cite{Kotliar_et_al:2006} to assess and analyze the different mechanisms proposed for the observed magnetostructural transition using a basis of trimer-centered Wannier orbitals. The larger spatial extent of such molecular trimer orbitals in comparison to more localized atomic-like orbitals leads to a non-negligible intertrimer Coulomb repulsion $V_{tr}$~\cite{Aretz_et_al:2025}. We hence consider both types of interactions, $U_{tr}$ and $V_{tr}$, and, in particular, focus on analyzing their influence on the physics of the bulk structure.

Our investigation of the different gap opening mechanism due to intra- and intercluster interactions shows a clear disagreement with the charge disproportionation picture~\cite{Haraguchi_et_al:2017, Haraguchi/Yoshimura:2024}. Instead, we obtain a continuous crossover between a Mott-insulating regime with strong antiferromagnetic intertrimer correlations and a band-insulator corresponding to the formation of intertrimer singlets~\cite{Sheckelton_et_al:2017, Gao_et_al:2023, Zhang_et_al:2023, Aretz_et_al:2025}. We conclude that the physics of \nbcl is well described within a picture of weakly coupled Hubbard dimers. Additionally, we also suggest a possible origin of the scissoring distortion reported in Sheckelton~\textit{et al.} \cite{Sheckelton_et_al:2017} resulting from atomic level interactions within the trimer. We corroborate our theoretical results with an experimental investigation of the structure and properties of \nbcl and \nbclbr, and suggest that synthesis conditions influence the LT phases of \nbclbrx materials.

%%%%%%%%%%%%%%%%%%%%%%%%%%%%%%%%%%%%%%%%%%%%%%%%%%%%%%%%%%%%%%%%%%%%%%%%%%%

\section{Results and discussion}

\subsection{Stability of the $R3$ charge ordered phase}

We begin by investigating the stability of the proposed trimer breathing mode coupling to charge order as described by Haraguchi~\textit{et al.}~\cite{Haraguchi_et_al:2017, Haraguchi/Yoshimura:2024}.

To do so, we construct molecular trimer Wannier functions from the $2a_1$ bands that lie close to the Fermi level~[\pref{fig:intro}(e)] following Ref.~\cite{Grytsiuk_et_al:2024}. We then perform a series of \dftuv calculations using these trimer orbitals as the basis, treating both the Coulomb interactions on one trimer ($U_{tr}$) and between adjacent trimers ($V_{tr}$). This is made possible by a recently developed extension of the \textsc{quantum espresso} code~\cite{Carta_et_al:2025}.

\begin{figure}
 \includegraphics[width=\columnwidth]{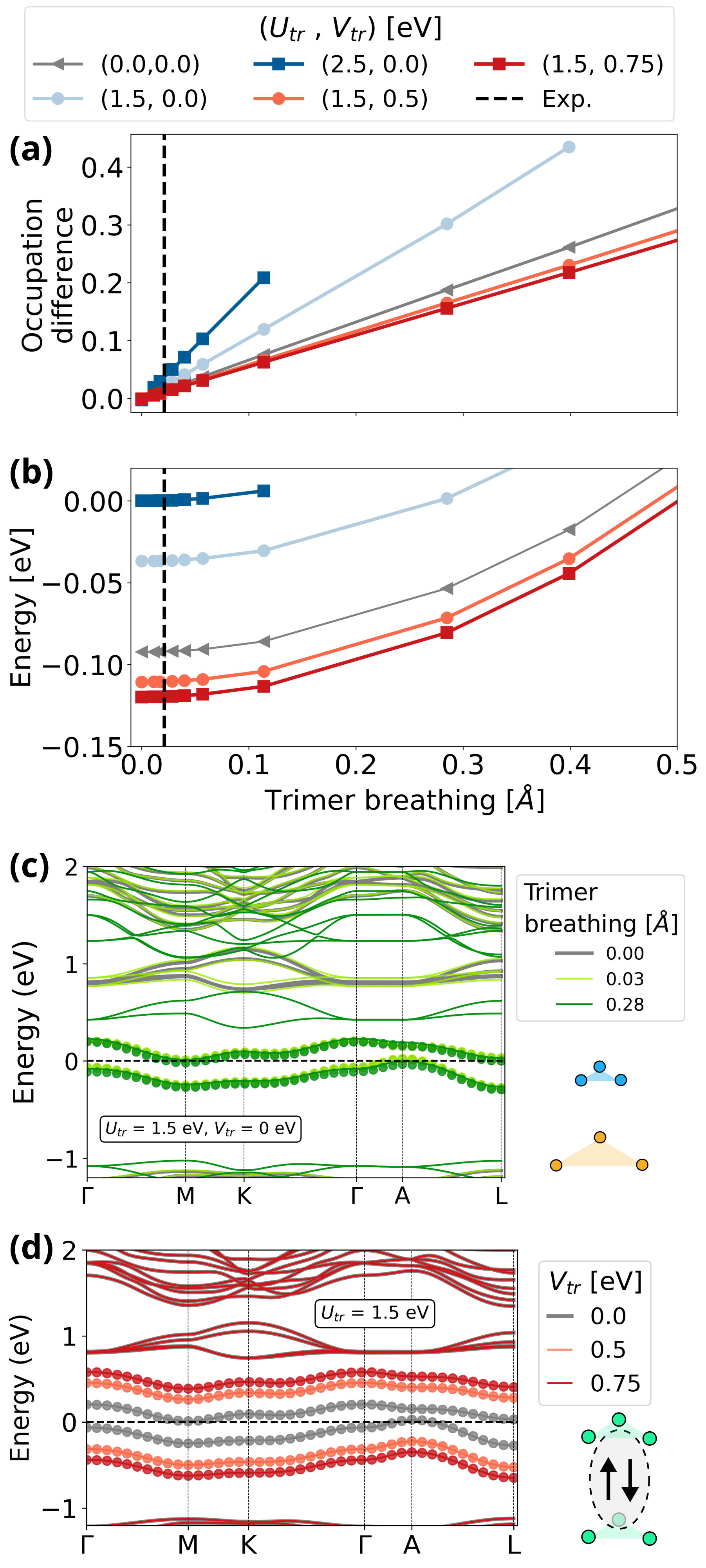}
 \caption{(a) Occupation difference between trimers and (b) total energy per unit cell as a function of the breathing mode for different values of $U_\text{tr}$ and $V_\text{tr}$. The vertical dashed line indicates the experimentally reported amplitude of the breathing mode from Ref.~\cite{Haraguchi_et_al:2017}. Band structure plots for (c) three breathing mode values at $(U_\text{tr}, V_\text{tr})=(1.5, 0)$\,eV and for (d) zero breathing mode ($R\bar{3}m$ structure) at three $V_\text{tr}$ values for $U_\text{tr}=1.5$ eV. The trimer bands are highlighted with full circles.}
 \label{fig:dftuv_haraguchi}
\end{figure}

The proposed breathing mode [see inset graphic in \pref{fig:dftuv_haraguchi}(c)] breaks the symmetry between the two trimers in the unit cell, leading to a difference in their onsite single-particle energies. Specifically, the energy of the long-edge trimer is lowered, while that of the short-edge trimer is raised.

In this section, we focus on spin unpolarized calculations, where the onsite occupation for spin up equals the onsite occupation for spin down, $n^\uparrow_{ii}=n^\downarrow_{ii}:=n_{ii}$. The spin polarized case will be treated in the next section. In this case, the Hubbard contribution to the Kohn-Sham potential due to the onsite interaction parameter reads:
\begin{equation}
\Delta v^\sigma_{ii} = U_\text{tr}\left(\frac{1}{2}-n_{ii}\right).
\label{eqn:U_dudarev}
\end{equation}
In the high-symmetry $R\bar{3}m$ structure, the two trimers are equivalent, leading to an onsite occupation of exactly $n_{ii} = 0.5$ for each one. Consequently, the potential correction from \pref{eqn:U_dudarev} is zero. A structural distortion, such as the trimer breathing mode that lowers the symmetry to $R3$, breaks this equivalence. This splitting of the trimer orbital energy levels creates a small initial occupation difference, $\Delta n$. The Hubbard $U$ term then acts to amplify this imbalance: for the more occupied trimer ($n_{ii} > 0.5$), $\Delta v^\sigma_{ii}$ is negative, favoring further filling, while for the less occupied one ($n_{ii} < 0.5$), $\Delta v^\sigma_{ii}$ is positive, promoting emptying.
Following the charge disproportionation model proposed by Haraguchi~\textit{et al.}~\cite{Haraguchi_et_al:2017}, a sufficiently strong Hubbard $U$ can be expected to drive this system towards a larger occupation difference, potentially approaching the limit of two electrons, thereby forming a singlet state on one trimer while completely emptying the other. However, our calculations reveal that this amplification effect is rather weak.

Our results are presented in \pref{fig:dftuv_haraguchi}(a, b) where we plot the occupation difference and the energy as a function of the trimer breathing mode amplitude which is defined as the edge difference between long-edge and short-edge trimers, for values around the calculated constrained random-phase approximation (cRPA) values~[$(U_\text{tr}, V_\text{tr}) = (1.46, 0.38)$\,eV, see \pref{tab:crpa_trimer}]. We remark that the absolute value of the energy difference between calculations with different $U_\text{tr}$ and $V_\text{tr}$ is not physically meaningful, and it is only shown in \pref{fig:dftuv_haraguchi}(b) for visual clarity.

We find that for $(U_\text{tr},V_\text{tr})=(1.5,0)$\,eV, the reported $R3$ structure~\cite{Haraguchi_et_al:2017, Haraguchi/Yoshimura:2024} corresponds to an occupation difference of only 0.03 electrons [see vertical dashed line in \pref{fig:dftuv_haraguchi}(a)]. We observe that even at $U_\text{tr} = 2.5$\,eV ($60\%$ higher than the computed cRPA value) and at a breathing mode amplitude of 0.4~\AA{} (20 times the experimental value), the occupation difference remains smaller than 0.4 electrons. Furthermore, we observe that the energy minimum consistently occurs at zero breathing mode, which corresponds to the higher symmetry $R\bar3m$ structure~[\pref{fig:dftuv_haraguchi}(b)].

By examining the band structure at different breathing mode amplitudes [\pref{fig:dftuv_haraguchi}(c)], we observe minimal changes in the bands near the Fermi level as the breathing mode amplitude increases, and the material remains metallic. This finding is not consistent with the experimentally observed insulating behavior.

To investigate this point further we perform calculations where in addition to the +$U$ correction we include an intersite term $V_\text{tr}$. The +$V$ correction essentially modifies the effective hopping between neighboring sites by adding an intersite potential~\cite{LeiriaCampoJr/Cococcioni:2010}:
\begin{equation}
 \Delta v^\sigma_{ij} = -V_\text{tr} n^\sigma_{ji} \quad .%{c_{j}^{\sigma}}^\dag c_{i}^{\sigma} \quad .
 \label{eqn:V_shift}
\end{equation}
Here $n^\sigma_{ji}$ is the off-diagonal intertrimer component of the density matrix which is a measure of the intersite hybridization. 

Upon including $V_\text{tr}$ [light red and red lines in \pref{fig:dftuv_haraguchi}(a, b)], we observe a further suppression of the occupation difference between trimers. This is accompanied by an opening of a band gap [\pref{fig:dftuv_haraguchi}(d)], the nature of which will be discussed more in detail in the next section.

Consequently, based on our \dftuv calculations, the charge-ordered phase accompanied by a trimer breathing mode as proposed by Haraguchi~\textit{et al.}~\cite{Haraguchi_et_al:2017} appears to be energetically unfavorable. We therefore rule out the charge disproportionation mechanism as explanation for the observed magnetic-nonmagnetic transition in \nbcl.

\subsection{Origin of the quenched magnetism in the $R\bar{3}m$ phase}

We now consider different possible explanations of the low susceptibility observed in $\beta$-\nbcl.
In this section, we study separately the effects of the interaction terms $U_\text{tr}$ and $V_\text{tr}$, allowing for spin-polarization in the system.

As previously mentioned, with no spin-polarization, one would expect the $+U$ correction to promote charge polarization in one of the trimer orbitals. In spin-polarized calculations instead, the $+U$ correction can favor the formation of an onsite magnetic moment on the Nb$_3$ molecular orbital. This is because the potential shift $\Delta v^\sigma_{ii} = U_\text{tr}\left(\frac{1}{2}-n^\sigma_{ii}\right)$ for site $i$ and spin $\sigma$, promotes either fully occupied or completely empty spin channels on each orbital. So, at strong $U$, one would expect for one trimer orbital one spin channel $\sigma$ to be completely occupied $n^\sigma_{ii}\sim1$ and the opposite spin channel $\bar{\sigma}$ to be empty $n^{\bar{\sigma}}_{ii} \sim 0$, which corresponds to the formation of a local magnetic moment on the trimer.

\begin{figure}[htb!]
  \includegraphics[width=\columnwidth]{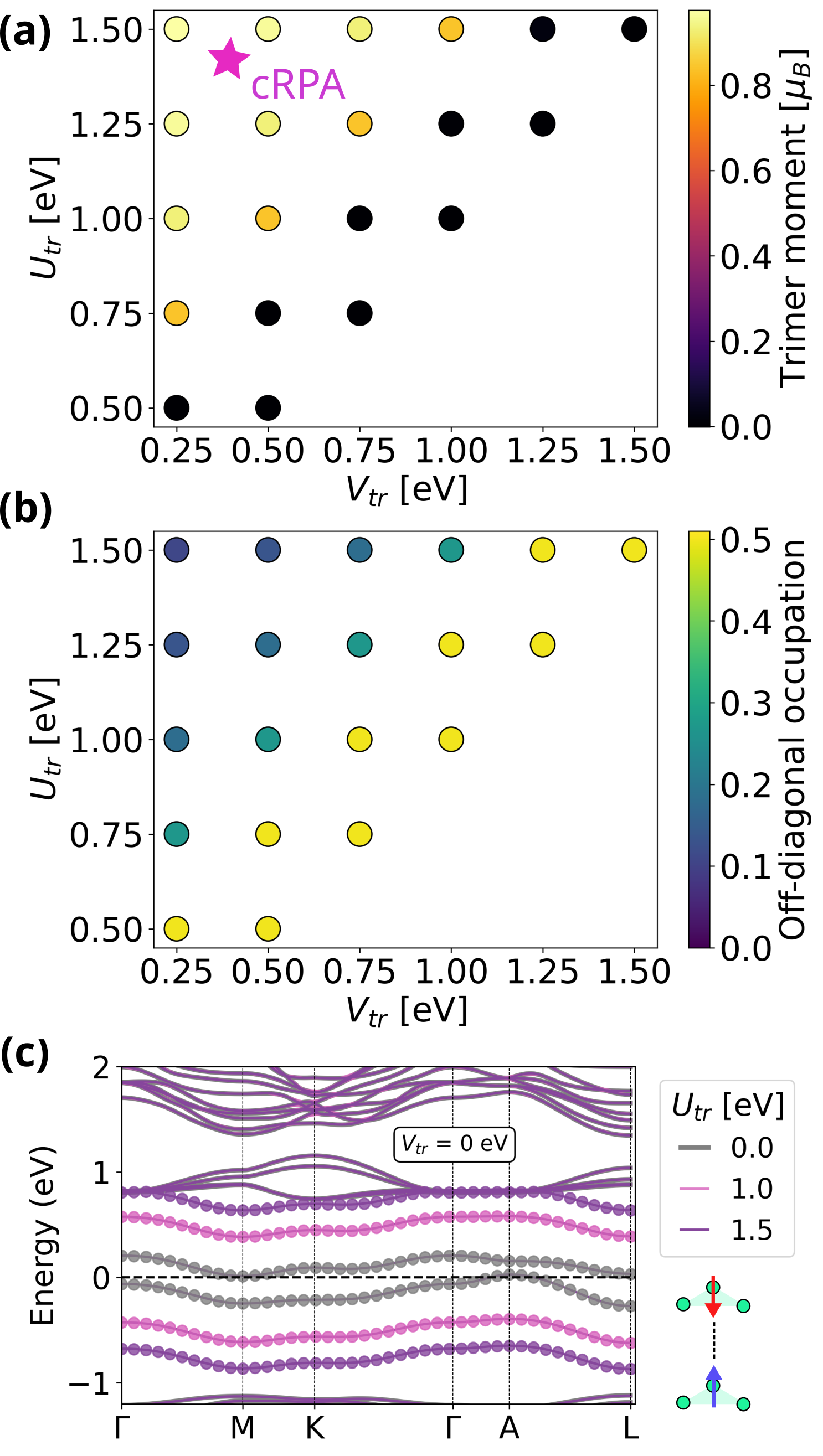}
  \caption{
  %\PM{larger markers for circles in a and b} \PM{switch order of d and c} 
   (a) Absolute value of the trimer moment and (b) intertrimer occupation as a function of $U_\text{tr}$ and $V_\text{tr}$. Star indicates the cRPA values. (c) Band structure within spin-polarized \dftuv with antiferromagnetically aligned trimer magnetic moments for three $U_\text{tr}$ values for $V_\text{tr}=0$. Bands corresponding to the trimer orbitals are highlighted with circles. All calculations are performed for the $R\bar{3}m$ structure (no trimer breathing mode).}
  \label{fig:dftuv_sheckleton}
\end{figure}

In \pref{fig:dftuv_sheckleton}(a, b), we show the average onsite trimer moment $|n^\uparrow_{ii}-n^\downarrow_{ii}|$ and the magnitude of the off-diagonal occupation matrix element $n^\sigma_{ij}$, respectively, as a function of $U_\text{tr}$ and $V_\text{tr}$. In \pref{fig:dftuv_sheckleton}(c), we show the band structure for fixed $V_\text{tr}$ and changing $U_\text{tr}$.% [the effect of varying $V_{tr}$ at fixed $U_{tr}$ was shown in \pref{fig:dftuv_haraguchi}(d) for the non-magnetic case].%, and vice versa, respectively.

Two competing tendencies emerge: a strong onsite Coulomb interaction $U_\text{tr}$ favors the development of antiferromagnetically aligned local magnetic moments on the Nb$_3$ sites~[\pref{fig:dftuv_sheckleton}(a)], suppressing any intersite bonding [$n^\sigma_{ij}\rightarrow0$ in \pref{fig:dftuv_sheckleton}(b)].

This local moment formation leads to a band gap opening~[\pref{fig:dftuv_sheckleton}(c)] and the strong antiferromagnetic correlations between adjacent trimers could be responsible for the suppression of the magnetic susceptibility~\cite{Grytsiuk_et_al:2024}. We underline that we are not arguing for the presence of long-range antiferromagnetic order in the material. Rather, we argue that in the strong $U_\text{tr}$ regime nearest neighboring trimer orbitals could show strong antiferromagnetic correlations, as schematically depicted in the inset in \pref{fig:dftuv_sheckleton}(c). Conversely, a dominant intersite Coulomb interaction $V_\text{tr}$ promotes the formation of bonding-antibonding states between trimers ($n^\sigma_{ij}=0.5$)~[\pref{fig:dftuv_sheckleton}(b)] and also leads to the formation of a band gap~[as was previously shown in \pref{fig:dftuv_haraguchi}(d)], which in this case is related to the large bonding-antibonding splitting. Since two electrons per dimer would then occupy the lower-lying bonding state, forming a nonmagnetic singlet [inset in \pref{fig:dftuv_haraguchi}(d)], this scenario would also be compatible with a drop in magnetic susceptibility.

Our cRPA calculations for the trimer molecular orbitals, in good agreement with the findings of Grytsiuk, Aretz~\textit{et al.}~\cite{Grytsiuk_et_al:2024,Aretz_et_al:2025}, suggest that \nbcl lies closer to the strong $U_\text{tr}$ regime.

To summarize, our calculations indicate that a change in stacking from $P\bar{3}m1$ to $R\bar{3}m$ is sufficient to explain the low magnetic susceptibility observed in $\beta$-\nbcl, without the need to introduce any further symmetry-lowering distortions, in agreement with previous works~\cite{Grytsiuk_et_al:2024, Aretz_et_al:2025}.

\subsection{Dynamical effects in the $R\bar{3}m$ phase}

So far, our treatment of the physics in \nbcl has been based on a static mean-field treatment of interaction effects, which highlights the crucial role of both onsite and intersite interactions, particularly within dimerized units of Nb$_3$ molecular orbitals. To move beyond this static mean-field approximation and capture dynamical effects, we now employ DFT+DMFT.

We construct a DMFT impurity problem for the $R\bar{3}m$ structure using a cluster of two neighboring molecular trimer orbitals, as first introduced in Aretz~\textit{et al.}~\cite{Aretz_et_al:2025}, and explore the behavior of the system under a variety of interaction parameters, in particular exploring the nature of the insulating state.

This approach is well suited for systems like \nbcl due to the strong onsite and intersite interactions $U_{tr}$ and $V_{tr}$, a significant intertrimer hopping along the stacking direction ($t_\text{LT}=0.13$\,eV), and comparatively weaker hybridization between the dimerized cluster and the rest of the Nb$_3$ molecular orbitals ($t_\text{in-plane}=0.02$\,eV). These considerations allow us to draw significant analogies between the physics of the \nbx family and the so-called \textit{Hubbard dimer}, a model system consisting of two interacting sites with one orbital each.
	
For two electrons in the Hubbard dimer model (\textit{i.e.} at half filling), the gap, $E_g$, in the excitation spectrum between the highest occupied and lowest unoccupied states of the Hubbard dimer, is dependent on the three parameters that define the model, the intersite hopping $t$, the onsite $U$ and the intersite $V$ interaction parameters~\cite{Pavarini_et_al:2017}:
\begin{equation}
	\label{eqn:gap}
	E_g = - 2t + V + \sqrt{(U-V)^2 + 16t^2}.
\end{equation}
In general the (unnormalized) ground state of such a model system $| \psi^\text{dimer}_0 \rangle$ can be written as:
\begin{equation}
	\begin{split}
		\label{eqn:grd_state}
		| \psi^\text{dimer}_0 \rangle = A & (| \uparrow,\,\downarrow \rangle - | \downarrow,\,\uparrow \rangle ) \\
		+  &(| \uparrow \downarrow,\,0 \rangle + | 0,\,\uparrow \downarrow 
		\rangle ),
	\end{split}
\end{equation}
where $A = 4t/[\sqrt{(U-V)^2 + 16t^2} - (U-V)]$. Here, $|\uparrow,\,\downarrow \rangle$ corresponds to a state with an up-spin electron on the first and a down spin electron on the second site, whereas $| \uparrow \downarrow,\,0 \rangle$ corresponds to double occupation of the first site, and analogous for the other states in \pref{eqn:grd_state}.
	
For $U=V$, $A=1$ the ground state can be described by a single Slater determinant expressed in the bonding-antibonding basis, and corresponds to the bonded state of two trimer sites. The opposite limit, $U \gg V$, $A \rightarrow \infty$, corresponds to the suppression of double occupancies and the maximally correlated state, the Heitler-London limit.

\begin{figure*}[htb]
  \includegraphics[width=0.9\textwidth]{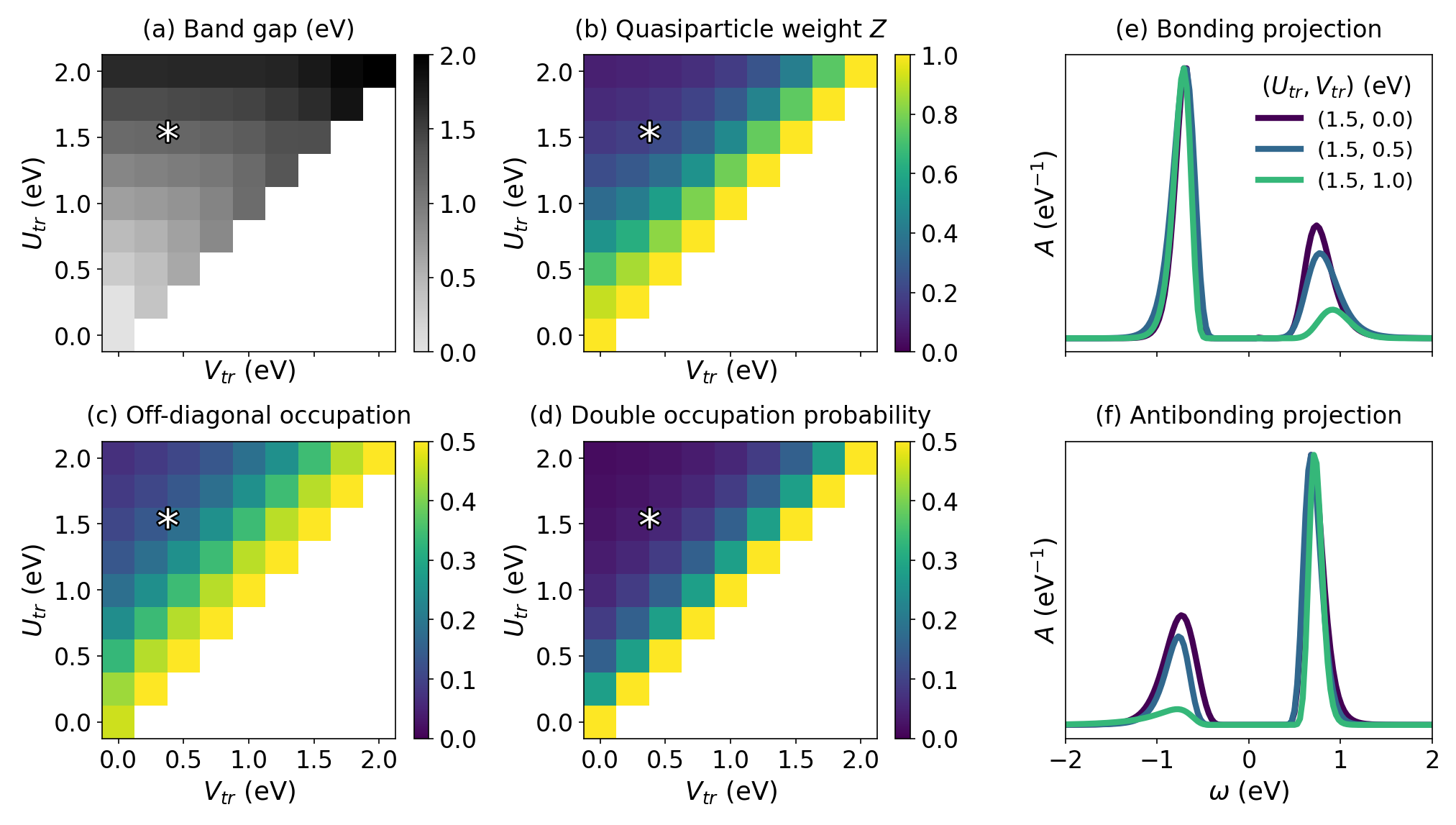}
  \caption{Observables of the $R\bar{3}m$ \nbcl cluster from DFT+DMFT as a function of $U_\text{tr}$ and $V_\text{tr}$. The cRPA values for \nbcl are shown with a star. (a) Band gap. (b) Quasiparticle weight, $Z$. (c) Off-diagonal occupation. (d) Double occupation probability. (e) Spectral function projected onto the bonding and (f) antibonding combination of the individual trimer orbitals across the ($U_\text{tr}$, $V_\text{tr}$) spectrum.} %\PM{try different colorscheme?}
  \label{fig:dmft}
\end{figure*}

In \pref{fig:dmft}(a-d), we depict the dependence of various properties of $\beta$-\nbcl on both $U_\text{tr}$ and $V_\text{tr}$. We show the band gap~[\pref{fig:dmft}(a)] obtained from the spectral function, and the quasiparticle weight~[\pref{fig:dmft}(b)] defined from the frequency dependence of the self-energy. In addition, we also show the off-diagonal occupation matrix element~[\pref{fig:dmft}(c)], and the probability of double occupancy on one of the Nb$_3$ molecular orbitals~[\pref{fig:dmft}(d)]. The latter is obtained from analyzing the probability of the various multiplet states occurring in the two site cluster, based on the statistics from the employed quantum Monte Carlo solver. 

We obtain an insulating solution for any value of the tested $U_\text{tr}$ and $V_\text{tr}$ parameters, except in the vicinity of $U_\text{tr}=V_\text{tr}=0$~[\pref{fig:dmft}(a)], with the size of the band gap depending on both $U_\text{tr}$ and $V_\text{tr}$. At high $U_\text{tr}$ and small $V_\text{tr}$, the insulator exhibits hallmark features of a Mott state, such as a vanishing quasiparticle weight~[\pref{fig:dmft}(b)]. By increasing $V_\text{tr}/U_\text{tr}$, the quasiparticle weight rises, reaching $Z = 1$ at $U_\text{tr} = V_\text{tr}$, where the system is a band insulator. This is accompanied by a gradual increase in the off-diagonal occupation from $V_\text{tr}=0$ to the maximum 0.5 at $U_\text{tr} = V_\text{tr}$, with the double occupation probability changing in a similar way~[\pref{fig:dmft}(c, d)]. At small $V_\text{tr}/U_\text{tr}$, our results show a marked suppression of double occupancies highlighting the localization of a single electron per Nb$_3$ site. The multiplet analysis can also clarify the relative spin orientation of the Nb$_3$ sites. Even at small $V_\text{tr}/U_\text{tr}$, configurations with spins aligning in opposite directions between trimers have high probability, while configurations where spins point in the same direction on both trimers are suppressed. 

The analogies to the Hubbard dimer are thereby particularly evident. We can identify the band insulator observed on the diagonal in \pref{fig:dmft}(a-d) with the $A=1$ bonding regime and the Mott insulator at low $V_\text{tr}/U_\text{tr}$ with the correlated high $A$ regime. Importantly, our results indicate a continuous crossover between the two different insulating states, which is reflected in all observables examined. 

Finally, we further exemplify this evolution by showing the spectral function at a fixed value of $U_\text{tr}$ and different values of $V_\text{tr}$, projected on suitably constructed bonding and antibonding dimer orbitals~[\pref{fig:dmft}(e, f)]. We observe that the two peaks change their character with increasing $V_\text{tr}$. The lower peak becomes predominantly bonding, while the upper peak becomes more antibonding as $V_\text{tr}$ increases. Due to inherent difficulties in the analytic continuation and the large size of the gap, the spectral function shows smooth peaks across the spectrum, and we are not able to resolve finer spectral details such as the splitting of the Hubbard peaks that was observed, e.g., in the work of Aretz~\textit{et al.}~\cite{Aretz_et_al:2025}.

These results provide evidence that the physics of the $R\bar{3}m$ phase of \nbcl and the \nbx family in general can be described by dimers of Nb$_3$ molecular units, which are weakly coupled to each other. These findings support the results from the previous sections and indicate the presence of strong local antiferromagnetic correlations which could be ultimately responsible for the suppression of the magnetic susceptibility in $\beta$-\nbcl, as has been suggested previously~\cite{Aretz_et_al:2025}. The additional analysis of the $U$ and $V$ dependence sheds light on the evolution of this state under stronger electronic correlation effects. Moreover, we explicitly demonstrate a continuous crossover between the Mott and singlet (band) insulating states, in line with recent theoretical proposals~\cite{Najera_et_al:2018, Zhang_et_al:2023, Jiang_et_al:2023}. The cRPA calculations indicate that \nbcl itself lies closer to the Mott limit, while the rest of the family lies closer to the singlet (band) insulating limit [see \pref{tab:crpa_trimer} and \pref{fig:dmft-nbx} for the rest of the family].

%%%%%%%%%%%%%%%%%%%%%%%%%%%%%%%%%%%%%%%%%%%%%%%%%%%%%%%%%%%%%%%%%%%%%%%%%%%%

\subsection{Further symmetry lowering in the $\beta$-\nbcl phase}

\begin{figure*}[htb]
 \includegraphics[width=0.9\textwidth]{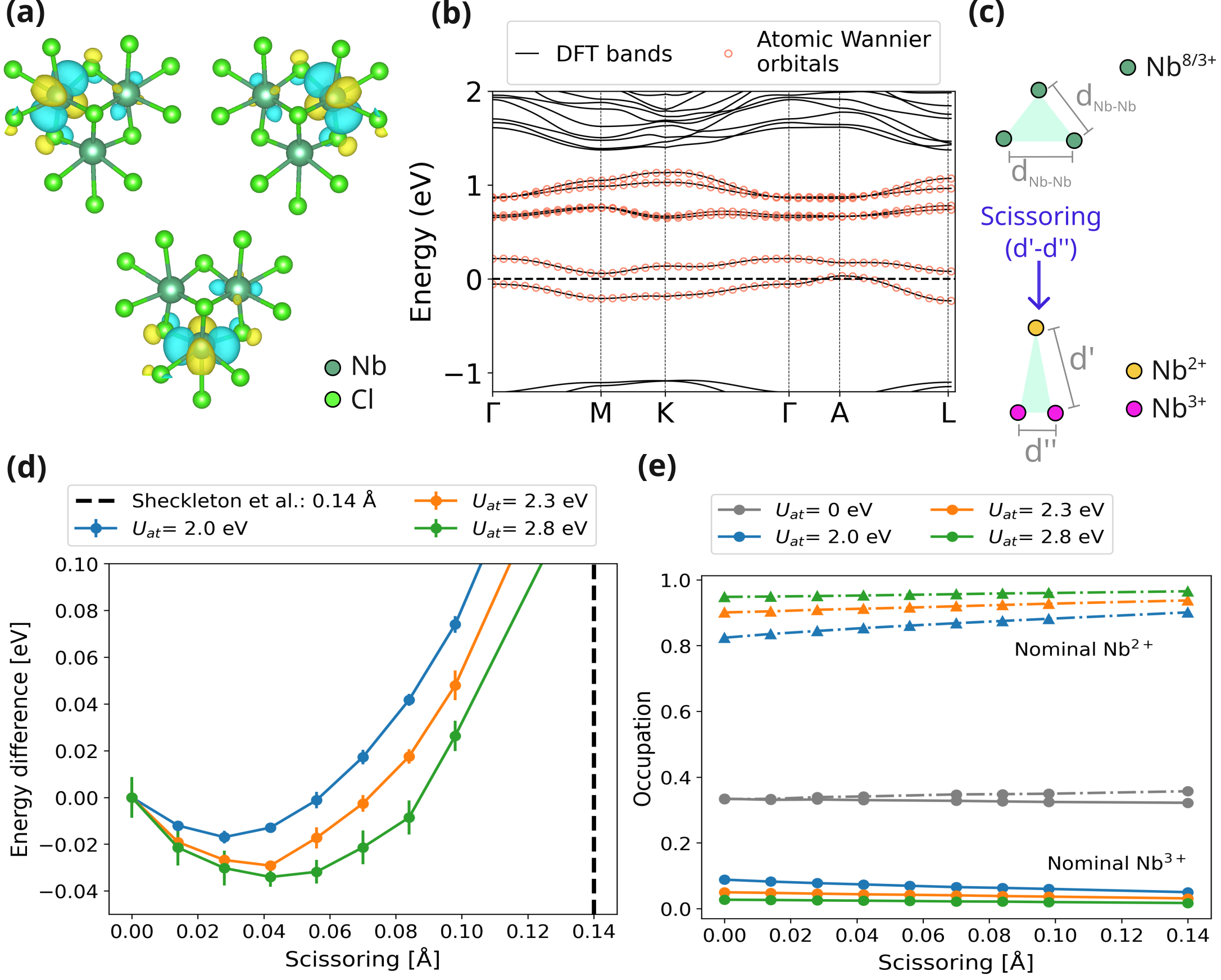}
 \caption{(a) Top view of the three atomic Wannier functions per trimer. (b) DFT band structure with the bands recalculated from the atomic Wannier model. (c) Schematic of the scissoring distortion within one trimer, connecting the $R\bar{3}m$ structure to the $C2/m$. (d) Energy difference per unit cell as a function of the scissoring amplitude. The black dashed line corresponds to the scissoring amplitude reported in Ref.~\cite{Sheckelton_et_al:2017}. (e) Occupation of the inequivalent Nb atoms inside one trimer. The solid lines correspond to the symmetry-equivalent nominal Nb$^{3+}$ sites, while the dash-dotted line corresponds to the nominal Nb$^{2+}$ site. }
 \label{fig:internal_disproportionation}
\end{figure*}

We will now consider the case of a system very close to the Mott regime, with relatively little intertrimer hybridization as a starting point to understand further symmetry lowering effects, analyzing the physics within the Nb$_3$ unit. By incorporating all six bands between 1.2\,eV and $-0.5$\,eV [as shown in Fig.~\ref{fig:internal_disproportionation}(b), corresponding to the $2a_1$ and $2e$ levels in Fig.~\ref{fig:intro}(d)] in the wannierization, we construct a basis set consisting of one $d$ atomic orbital per Nb atom, yielding a total of three $d$ orbitals per Nb$_3$ trimer [as shown in \pref{fig:internal_disproportionation}(a)] and six $d$ orbitals per unit cell~\cite{Streltsov_et_al:2025}. 

Using cRPA, we calculated the relevant, now atomic, interaction parameters [\pref{tab:crpa_atom}]: the onsite interaction on a single Nb atom, $U_\text{at}=2.3$\,eV, the intratrimer interaction between two different Nb atoms inside the same trimer, $V^\text{intra}_\text{at}=1.36$\,eV, and the intertrimer interaction between neighboring trimers, $V^\text{inter}_\text{at}=0.39$\,eV.
	
Leveraging our previous finding that the intertrimer occupation is small in the Mott regime, leading to electron localization on the trimer, we neglect the interaction across trimers ($V^\text{inter}_\text{at}$) for simplicity. We then perform DFT+DMFT calculations considering two separate impurity problems, with each impurity containing the three Nb $d$ orbitals, including $U_\text{at}$ and $V^\text{intra}_\text{at}$ in the interaction Hamiltonian [see \pref{eqn:h_int_at} for further details].

While DFT calculations distribute the single electron per each trimer equally among the Nb atoms, leading to an oxidation state of Nb$^{8/3+}$, in our charge self-consistent DFT+DMFT calculations we observe that for realistic values of $(U_\text{at}, V^\text{intra}_\text{at})$ most of the charge concentrates on one of the Nb atoms~[see \pref{fig:internal_disproportionation}(e) for 0.0~\AA{} scissoring]. This charge ordering within the trimer changes the nominal oxidation states of the three Nb atoms and renders them inequivalent: for every Nb$_3$ trimer we obtain one nominal Nb$^{2+}$ site and two empty Nb$^{3+}$ sites.

Our results also indicate that the structure accommodates this change in oxidation state within each trimer by elongating the bond length $d'$ between the nominal Nb$^{3+}$ and Nb$^{2+}$ atoms and contracting the bond length $d''$ between the two Nb$^{3+}$ atoms~[see \pref{fig:internal_disproportionation}(c)]. The difference, $d'-d''$, is termed \textit{scissoring}. This scissoring distortion lowers the symmetry from $R\bar{3}m$ to $C2/m$, precisely matching both our own (see next section) and previous experimental work by Sheckelton~\textit{et al.}~\cite{Sheckelton_et_al:2017}.

This can be seen from Fig.~\ref{fig:internal_disproportionation}(d, e), where we show the energy per unit cell and the occupation of the nominal Nb$^{2+}$ and Nb$^{3+}$ as a function of the scissoring mode. We consider various values of $U_\text{at}$, while maintaining a constant ratio of $V^\text{intra}_\text{at}/U_\text{at}=0.24$ (corresponding to our cRPA results). However, it is worth noting that our results exhibit weak dependence on the value of $V^\text{intra}_\text{at}$, and similar outcomes can be obtained by setting $V^\text{intra}_\text{at}=0$.

For $U_\text{at}=0$\,eV, the scissoring changes the occupation of the Nb atoms only slightly [\pref{fig:internal_disproportionation}(e), gray line], and the system always relaxes back to $R\bar{3}m$ symmetry. For all finite values of $U_\text{at}$ considered, the system shows strong electron charge concentration on the nominal Nb$^{2+}$ atom~[\pref{fig:internal_disproportionation}(e), colored lines], and an energy minimum at a finite scissoring distortion~[\pref{fig:internal_disproportionation}(d)]. As $U_\text{at}$ increases, the energy minimum deepens and shifts towards a larger scissoring amplitude, while the occupation of the Nb$^{2+}$ atom becomes closer to 1. 

Compared to the results of Sheckelton~\textit{et al.}~\cite{Sheckelton_et_al:2017}, our calculated equilibrium scissoring amplitude is markedly smaller. For $U_\text{at}=2.3$ eV [orange solid line in \pref{fig:internal_disproportionation}(d)], we find $\sim$0.04~Å, versus 0.14~Å reported in Ref.~\cite{Sheckelton_et_al:2017} [indicated by the vertical black line in \pref{fig:internal_disproportionation}(d)]. 

We also note that comparable results can be obtained through simpler DFT+$U$ calculations, utilizing standard orthonormalized atomic orbitals as implemented in \textsc{quantum espresso}~\cite{Giannozzi_et_al:2017}. If one starts an atomic relaxation with a small scissoring distortion and initializes the entire magnetic moment on a single Nb atom per Nb$_3$ trimer (and ensuring that the two magnetic Nb atoms in a unit cell containing two trimers are initialized with opposite magnetization), then for realistic values of the Hubbard $U$ parameter, the system relaxes to $C2/m$ symmetry. This $C2/m$ phase is lower in energy than the antiferromagnetic $R\bar{3}m$ phase, where the magnetization is uniformly spread over the trimer, and it exhibits a scissoring mode amplitude similar to that found in our DFT+DMFT calculations, with most of the trimer's magnetization concentrated on the nominal Nb$^{2+}$ atom.

In this section, we have demonstrated the possibility of additional ordering in $\beta$-\nbcl arising from the internal structure of the Nb$_3$ unit. Since the amplitude of the scissoring distortion is dependent on the value of $U_\text{at}$, and given that intertrimer bonding components become more relevant for other members of the \nbx family, the approximation of being deep in the Mott state may break down in those cases. We would then expect this scissoring tendency to be less pronounced across the rest of the series, which aligns with experimental observations, having previously found the $C2/m$ phase only in \nbcl. Hence, we now turn to our experimental results for \nbcl and the mixed \nbclbr compound.

%%%%%%%%%%%%%%%%%%%%%%%%%%%%%%%%%%%%%%%%%%%%%%%%%%%%%%%%%%%%%%%%%%%%%%%%%%%%

\subsection{Experimental}
\label{sec:experiment}

\textbf{Crystallography.} In order to validate the results of our computational investigation of \nbcl, and to attempt to understand the origins of the differing LT phases reported in the literature, we have synthesized \nbcl and \nbclbr and characterized their crystallographic structure and electronic properties. Our samples of \nbcl were determined by single crystal XRD to adopt the HT $P\bar{3}m1$ phase at 100~K, the lowest temperature at which we could perform these measurements. Magnetic susceptibility measurements (\pref{fig:magSC}) on a single crystal show the phase transition to occur below 100 K on cooling and slightly above 100 K on heating, and our finding of the HT phase at 100 K is therefore ascribed to hysteresis of the phase transition.

\begin{figure}[htb]
  \includegraphics[width=0.95\columnwidth]{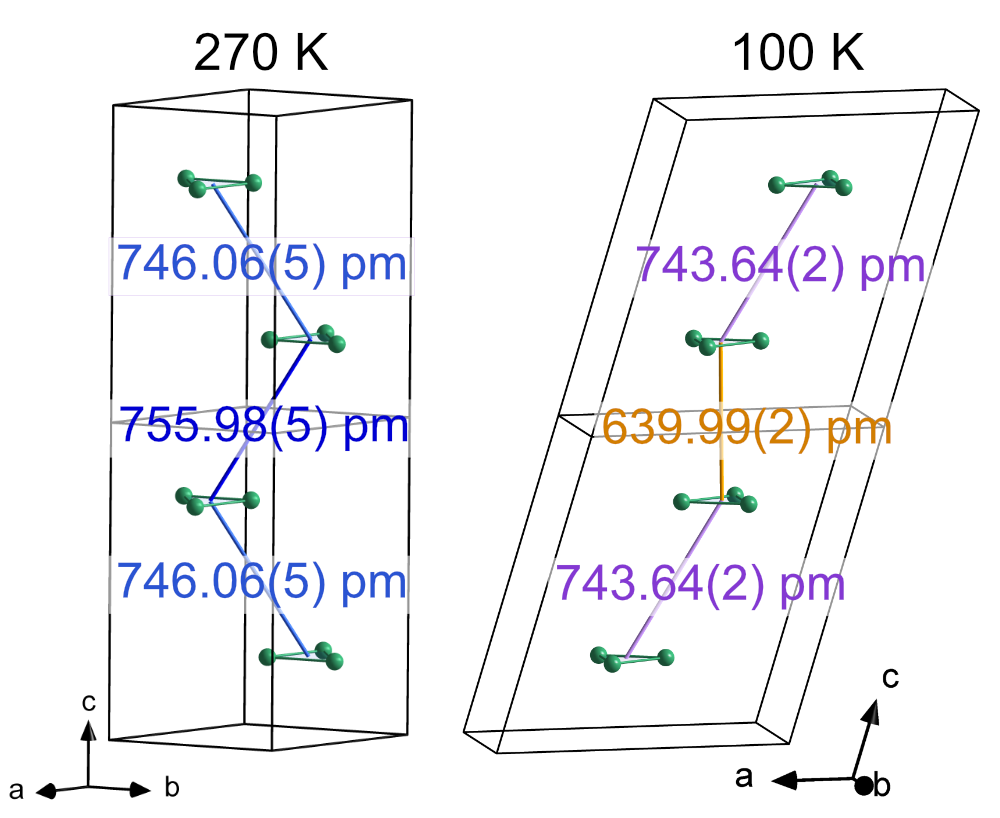}
  \caption{Stacking arrangement and distances of Nb$_3$ clusters (green triangles) in Nb$_3$Br$_{4-x}$Cl$_{4+x}$ as determined from single crystal X-ray diffraction data at 270\,K (left) and 100\,K (right). Chloride and bromide ions are omitted for clarity.}
  \label{fig:interclusterdistance}
\end{figure}

Therefore, in order to observe the LT phase, we prepared \nbclbrx, which was determined to have an increased phase transition temperature by Pasco~\textit{et al.}~\cite{Pasco_et_al:2019}. However, whereas the authors observed a LT $R\bar{3}m$ phase, we found the LT structure to adopt $C2/m$ (\pref{tab:crystdata}, \pref{fig:interclusterdistance}), indicating that the space group of LT \nbclbr, like that of LT \nbcl, possesses some ambiguity. 

\begin{table}
	\caption{Selected crystallographic data of Nb$_3$Br$_{4-x}$Cl$_{4+x}$ above and below the transition temperature.}
	\begin{tabular}{m{2.2cm}m{2.3cm}m{2.5cm}}
		\hline
		\hline
		Formula$^{*}$ & Nb$_3$Br$_{3.7}$Cl$_{4.3}$  & Nb$_3$Br$_{3.8}$Cl$_{4.2}$ \\ \hline
		CCDC \linebreak number & 2364410 & 2365009 \\
            Temperature & 270.0(1)\,K & 100.0(1)\,K \\
        Crystal \linebreak system & trigonal & monoclinic \\
        Space group & $P\bar{3}m$ & $C2/m$\\
        Wavelength & 154.184\,pm & 71.037\,pm \\
        Radiation type & Cu-K$_\alpha$ & Mo-K$_\alpha$ \\
        Residual factor ($R_1$) & 0.0350 & 0.0350 \\
        Residual factor ($wR_2$) & 0.0903 & 0.0909 \\
         \hline
        \multicolumn{3}{p{0.9\columnwidth}}{\begin{footnotesize}
            $^{*}$The composition variation is within the synthesis and refinement error margin, and is considered as Nb$_3$Br$_{4}$Cl$_{4}$.
        \end{footnotesize}}\\
		\hline
		\hline
	\end{tabular}
	\label{tab:crystdata}
\end{table}

Notably, our synthetic method, which employed only elemental Nb, NbCl$_5$ and NbBr$_5$ as reagents, follows that of Sheckelton~\textit{et al.}~\cite{Sheckelton_et_al:2017}, who found the $C2/m$ phase in \nbcl at low temperatures. In contrast, synthesis using a transport agent such as NH$_4$Cl or TeCl$_4$ were reported to yield a LT $R\bar{3}m$ phase~\cite{Pasco_et_al:2019, Kim_et_al:2023}, whereas reports of the $R3$ phase are associated with flux growth in PbCl$_3$ followed by soaking in hot water~\cite{Haraguchi_et_al:2017, Haraguchi/Yoshimura:2024}. These associations suggest that the precise LT phase adopted by \nbcl and \nbclbrx could be determined by the synthetic conditions. In particular, tellurium has been reported to incorporate itself into the \nbcl structure by capping Nb$_3$ triangles~\cite{Miller:1995}, and the introduction of defects has been suggested to trap Nb$_3$Br$_8$ in its HT phase~\cite{Pasco_et_al:2019}; a similar effect related to the introduction of impurities from the transport agent could push the material into the $R\bar{3}m$ phase, which is more closely related to the HT $P\bar{3}m1$ phase than the $C2/m$ is.

The changes between trimers in adjacent layers in \nbclbr following the phase transition can be readily seen in the single-crystal XRD refinements. At 270\,K, the intertrimer distances are at 755.98(5)\,pm and 746.06(5)\,pm (\pref{fig:interclusterdistance}, left). In the low temperature structure (at 100\,K), these distances alternate between 639.99(2)\,pm and 743.64(2)\,pm (\pref{fig:interclusterdistance}, right), leading to the stacking shift (\pref{fig:intro}b) which modifies the intercluster interactions (\pref{fig:intro}c).

\begin{figure}[htb]
  \includegraphics[width=0.85\columnwidth]{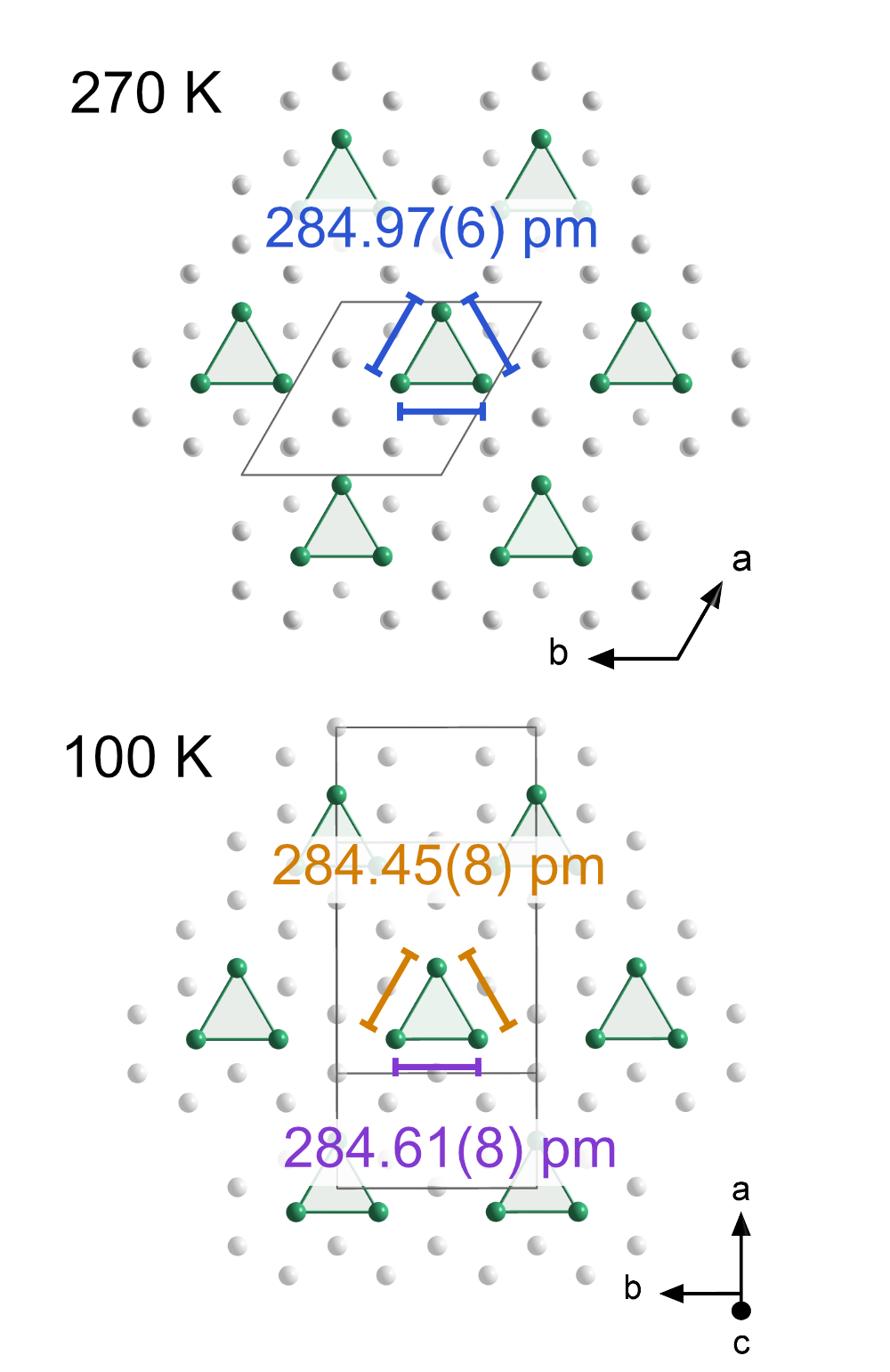}
  \caption{Nb--Nb distances within Nb$_3$ clusters in Nb$_3$Br$_{4-x}$Cl$_{4+x}$ as determined from single crystal X-ray diffraction data at 270\,K (top) and 100\,K (bottom). Niobium atoms are depicted in green, chloride and bromide ions are gray. The unit cell vectors are shown as grey lines.}
  \label{fig:intraclusterdistance}
\end{figure}

At 270~K, in the $P\bar{3}m1$ phase, the Nb--Nb distances within a cluster form equilateral triangles with bond lengths of 284.97(6)~pm (\pref{fig:intraclusterdistance}, top). In the low-temperature $C2/m$ phase at 100~K, the clusters adopt an isosceles triangular arrangement, with one Nb--Nb bond measuring 284.61(8)~pm and the other two at 284.45(8)~pm (\pref{fig:intraclusterdistance}, bottom). The asymmetry arises due to the unambiguous loss of 3-fold rotational symmetry upon cooling, but the overall change in bond lengths remains very small, to the point where one short and two long Nb--Nb bonds is possible within the experimental uncertainty. The difference in Nb--Nb bond lengths is much smaller in \nbclbr than was reported by Sheckelton~\textit{et al.} in \nbcl~\cite{Sheckelton_et_al:2017}. According to our computational results discussed in the previous section, the change in the stacking sequence observed in \pref{fig:interclusterdistance} is driven by intertrimer interaction, while the symmetry reduction to $C2/m$, which removes the 3-fold rotation axis and allows the Nb--Nb bond lengths to differ, is likely a secondary effect driven by intratrimer interactions. 

\textbf{Magnetic measurements.} We conducted magnetic measurements on \nbcl to observe the phase transition, as this occured below the minumum temperature at which we could perfom crystallographic measurements. We see a similar phase transition temperature (\textit{ca.} 100~K) and degree of hysteresis (\pref{fig:magSC}) to previous reports of single crystals of \nbcl~\cite{Haraguchi_et_al:2017, Sheckelton_et_al:2017, Kim_et_al:2023}. However, unlike Haraguchi~\textit{et al.}~\cite{Haraguchi_et_al:2017}, who observed this transition only in single crystals, we were able to detect a magnetic phase transition in both single crystal and powder samples (\pref{fig:magpowder}). In comparison to the single crystal sample, the powder sample displays an increased hysteresis, and a stronger temperature dependence of the paramagnetic response below the transition temperature. This temperature dependence leads to suppression of the paramagnetism near the transition temperature upon heating, a phenomenon also observed in assemblages of single crystals~\cite{Pasco_et_al:2019}, suggesting that inter-grain interactions affect the magnetic properties of \nbcl.

\textbf{Optical Band Gaps.} Optical measurements were performed above the structural transition temperature at room temperature. For both \nbcl and \nbclbr, we observe similar direct and indirect optical band gaps (see \pref{tab:optbandgaps}), based on measurements on bulk samples consisting of small crystallites. This result indicates similar electronic properties of \nbcl and \nbclbr, validating our use of \nbclbr as a proxy for low-temperature structural measurements. The values are consistent with literature reports~\cite{Date_et_al:2025,Yoon_et_al:2020} but are smaller compared to our DFT+DFMT results.

\begin{table}
	\caption{Direct and indirect optical band gaps for \nbcl and Nb$_3$Br$_4$Cl$_4$ at room temperature, determined by DRIFT Spectroscopy. (See \pref{fig:optbandnbcl} and \pref{fig:optbandnbbrcl} for full optical spectra)}
    \label{tab:optbandgaps}
	\begin{tabular}{m{2cm}m{2cm}m{2cm}}
		\hline
		\hline
		\multirow{2}{*}{\begin{tabular}[c]{@{}l@{}} \nbcl \end{tabular}} 
		& direct   & 0.510\,eV \\
		& indirect      &  0.344\,eV  \\[0.2cm]
		\multirow{2}{*}{\begin{tabular}[c]{@{}l@{}} Nb$_3$Br$_4$Cl$_4$ \end{tabular}} 
		& direct   &   0.537\,eV \\
		& indirect   &    0.311\,eV \\
		\hline
		\hline
	\end{tabular}
\end{table}

\textbf{Electrical conductivity.} We investigated the temperature-dependent conductivity of bulk crystals and observed semiconducting behavior (\pref{fig:elcond}) for the high temperature modification above 100\,K. The conductivity ranges from approximately $2 \times 10^{-3}$ S$\cdot$m$^{-1}$ at 300 K to $8 \times 10^{-8}$ S$\cdot$m$^{-1}$ at 100\,K. Below 100 K, the conductivity fell below the detection limit. No hysteresis was observed during cooling and heating cycles, which is consistent with the measurements being conducted above the structural transition temperature. As the magnetic susceptibility shows some hysteresis above 100 K (\pref{fig:magSC}), our experimental results suggest that the electrical conductivity is not strongly affected by the magnetic transition, further indicating that the magnetic transition is not driven by charge disproportionation. Our results algin with the earlier measurements by Yoon~\textit{et al.}~\cite{Yoon_et_al:2020} at higher temperatures.

%%%%%%%%%%%%%%%%%%%%%%%%%%%%%%%%%%%%%%%%%%%%%%%%%%%%%%%%%%%%%%%%%%%%%%%%%%%%

\section{Conclusion}

In this work, we performed a \dftuv, DFT+DMFT, and experimental study of the bulk LT $\beta$-\nbcl phase. Particularly, we studied the different proposed mechanisms of the observed magnetostructural transition, with a focus on the effects of intra- and intertrimer Coulomb interactions.

We showed that it is the dimerization of the layers due to the change in stacking that is the primary driver of the magnetic transition in these compounds and that the charge disproportionation model~\cite{Haraguchi_et_al:2017, Haraguchi/Yoshimura:2024} is not consistent with our calculations. In particular, through a DFT+DMFT analysis we show that \nbcl behaves as a system of weakly-coupled Hubbard dimers, and that two of the proposed explanations for the physics of \nbcl, the Mott-like strongly antiferromagnetically coupled system, and the singlet-like band insulator are smoothly connected, with the intertrimer interaction dictating their crossover. Additionally, the analysis of the intratrimer behavior suggests a possible mechanism for the experimentally observed scissoring of the trimer units.

The important role of the intertrimer interaction manifests itself also in the other members of the \nbx family~[see \pref{tab:crpa_trimer}] where a smaller $U/t$ ratio leads to a larger proximity to the band insulator solution rather than a Mott-like behavior.

In line with the theoretical findings, single-crystal X-ray diffraction reveals a structural phase transition in Nb$_3$Br$_{4}$Cl$_{4}$ from $P\bar{3}m1$ to $C2/m$, accompanied by a pronounced reduction in the distances between adjacent clusters. The bond lengths within the Nb$_3$ clusters slightly decrease upon cooling across the transition. Symmetry breaking leads to a slight distortion of the equilateral Nb$_3$ triangles into isosceles ones. However, the difference between the shorter and longer Nb--Nb bond lengths remains small and experimentally does not indicate a clear preference regarding whether the one inequivalent bond is longer or shorter than the others. The primary structural effect detected by our X-ray diffraction study is therefore a rearrangement of clusters of adjacent layers relative to one another, rather than significant changes within the clusters themselves. Magnetic measurements on \nbcl confirm the expected magnetic-to-nonmagnetic transition around 90\,K, with clear hysteresis observed in both single crystals and powder samples. Temperature-dependent conductivity measurements showed a gradual decrease upon cooling to the transition temperature, without significant hysteresis. Below 100\,K, no current could be detected within the sensitivity of the measurement setup. The differing hysteresis behavior of the magnetic and electric properties is consistent with our computational results, in particular the absence of interlayer charge disproportionation.

In conclusion, our work provides further evidence that the magnetostructural transition in $\beta$-\nbcl is driven primarily by a change in crystal stacking to the $R\bar{3}m$ phase. The physics of this interlayer dimerization can be understood in terms of weakly-coupled Hubbard dimers formed from pairs of Nb trimers. The subtle competition between intratrimer and intertrimer interactions tunes the system across a crossover between the limits of a Mott insulator and a band insulator. Concurrently, Coulomb interactions within a trimer drive a secondary, more subtle scissoring distortion within the Nb$_3$ trimer units.

This mechanism highlights the general importance of interlayer coupling in determining the electronic and magnetic properties of the \nbx family of materials. Our calculations suggest that targeted control of external parameters such as pressure, chemical substitution, or strain, might allow systematic exploration of the crossover from Mott to band insulator and engineering of new functionalities in this and related cluster Mott systems.

%%%%%%%%%%%%%%%%%%%%%%%%%%%%%%%%%%%%%%%%%%%%%%%%%%%%%%%%%%%%%%%%%%%%%
%% The "Acknowledgement" section can be given in all manuscript
%% classes.  This should be given within the "acknowledgement"
%% environment, which will make the correct section or running title.
%%%%%%%%%%%%%%%%%%%%%%%%%%%%%%%%%%%%%%%%%%%%%%%%%%%%%%%%%%%%%%%%%%%%%
\begin{acknowledgement}

The authors thank S. Grytsiuk, J. Aretz, M. R\"osner, and M. I. Katsnelson for valuable discussions and initial input files. P.M. and N.A.S. acknowledge funding from the Swiss National Science Foundation (Grant number 209454) and from ETH Z\"urich. Calculations were performed on the ``Euler'' cluster of ETH Z\"urich and the Swiss National Supercomputing Center Eiger cluster under Project ID s1304. C.P.R. acknowledges support from the project FerrMion of the Ministry of Education, Youth and Sports, Czech Republic, co-funded by the European Union (CZ.02.01.01/00/22\_008/0004591). We would like to thank Dr. F. Strauß for conducting and analyzing the conductivity measurements and Prof. M. Scheele for providing access to the necessary equipment. We also thank Dr. J. Glaser for conducting and analyzing the magnetic measurements.

\end{acknowledgement}

%%%%%%%%%%%%%%%%%%%%%%%%%%%%%%%%%%%%%%%%%%%%%%%%%%%%%%%%%%%%%%%%%%%%%
%% The same is true for Supporting Information, which should use the
%% suppinfo environment.
%%%%%%%%%%%%%%%%%%%%%%%%%%%%%%%%%%%%%%%%%%%%%%%%%%%%%%%%%%%%%%%%%%%%%
\begin{suppinfo}

The Supporting Information for this article is available at ...

\end{suppinfo}

\clearpage

\section{Supporting Information}

\setcounter{equation}{0}
\setcounter{figure}{0}
\setcounter{table}{0}
\setcounter{section}{0}
\setcounter{page}{1}
\makeatletter
\renewcommand{\theequation}{S\arabic{equation}}
\renewcommand{\thefigure}{S\arabic{figure}}
\renewcommand{\thetable}{S\arabic{table}}

\section{Theoretical and computational details}

\subsection{DFT+\textit{U}+\textit{V} calculations}

We perform \dftuv calculations using the \textsc{quantum espresso} package~(v6.6)~\cite{Giannozzi_et_al:2009, Giannozzi_et_al:2017} within the generalized gradient approximation with the Perdew-Burke-Ernzerhof~\cite{Perdew/Burke/Ernzerhof:1996} exchange-correlation functional, utilizing the semiempirical DFT-D3 correction~\cite{Grimme_et_al:2010} to account for the van der Waals effects. We use the scalar-relativistic ultrasoft pseudopotentials from the GBRV library~\cite{Garrity_et_al:2014} with semicore 4$s$ and 4$p$ states included as valence for the Nb atoms.

We construct a localized set of orbitals using \textsc{wannier90}~(v3.1.0)~\cite{Mostofi_et_al:2008, Pizzi_et_al:2020}. Starting from a conventional non-spin-polarized DFT calculations, we consider the two bands in close proximity to the Fermi level with dominant Nb contributions and with no entanglement with the surrounding bands~[see \pref{fig:intro}(e)]. Following Grytsiuk~\textit{et al.}~\cite{Grytsiuk_et_al:2024}, our basis set consists of two trimer-centered Wannier functions computed by projecting the Kohn-Sham (KS) states around the Fermi level onto initial guesses of $d_{z^2}$ orbitals and performing subsequent orthonormalization to finally obtain the molecular trimer orbitals shown in \pref{fig:intro}(f).

We then employ these trimer orbitals in both spin-unpolarized and spin-polarized \dftuv calculations using the implementation from Ref.~\cite{Carta_et_al:2025}. We consider both onsite (intratrimer) and intersite (intertrimer) interactions described by parameters $U_\text{tr}$ and $V_\text{tr}$, respectively [see \pref{fig:intro}(c)].

\subsection{DFT+DMFT calculations}

The trimer basis set for the DFT+DFMT calculations is described by the same Wannier functions obtained in the previous section.
We consider an impurity problem composed of two adjacent trimer orbitals, a ``dimer of trimers", for which the Hamiltonian reads:
\begin{equation}
H = H_0 + H_\text{int} - H_\text{DC},
\label{eqn:dmft_ham}
\end{equation}  
where $H_0$ corresponds to the effective single-particle DFT Hamiltonian as obtained in the wannierization procedure and $H_\text{int}$ is the interaction Hamiltonian: 
\begin{equation}
\label{eqn:h_int}
H_\text{int} = U_\text{tr} \sum_{i} \hat{n}_{i}^{\uparrow} \hat{n}_{i}^{\downarrow} 
+ V_\text{tr} \sum_{\langle ij \rangle, \sigma\sigma'} \hat{n}_{i}^{\sigma} \hat{n}_{j}^{\sigma'},
\end{equation}
where $\hat{n}_{i}^{\sigma}$ is the spin resolved number operator for trimer site $i$ and spin $\sigma$. The meaning of $U_{tr}$ and $V_{tr}$ corresponds to the same as in the \dftuv case, that of an onsite (intratrimer) and intersite (intertrimer) interaction, respectively.

$H_\text{DC}$ is the double-counting (DC) term, which cancels out the part of local interaction already included in the $H_0$ term~\cite{LeiriaCampoJr/Cococcioni:2010}.
Here, we employ the standard fully-localized limit for the expression of the energy of the double counting correction~\cite{Anisimov_et_al:1993}, which in this basis set, translates to a constant shift of the trimer orbital energies:
\begin{equation}
\label{eqn:dc_sigma}
\Sigma^{tr}_\text{DC} =  U  \sum_{i, \sigma} \bigg( N^\text{tr}_i -\frac{1}{2} \bigg)  \hat{n}_i^\sigma .
\end{equation}
Here, $N^\text{tr}_i = \sum_\sigma \langle n^\sigma_i \rangle $ is the occupation of the trimer orbitals.
In the presence of a difference in occupation between the two trimers, this term could in principle favor the localization of two electrons on one trimer.
However, similarly to \dftuv calculations, we find no trace of this tendency in our DFT+DMFT calculations and both trimers remain equally occupied.
We note that we have not included an intersite contribution to the double-counting potential. The approach outlined in Ref.~\cite{LeiriaCampoJr/Cococcioni:2010} and used in the \dftuv calculations, effectively removes the double counting of the Hartree contribution stemming from intersite interactions, which would further suppress charge-disproportionation in the present case. However, given that our DFT+DMFT calculations already show no tendency for charge order and maintain equally occupied trimers, we do not anticipate that including this term would significantly alter our results.

In the atomic basis, we construct Wannier functions including all 6 bands in the energy range between $-$0.5~eV and 1.5~eV around the Fermi level. This results in one $d$ orbital per Nb site oriented towards the center of the trimer.
We then perform DFT+DMFT calculations with 2 impurity models, each one including the 3 atomic orbitals belonging to one Nb$_3$ unit. The resulting Hamiltonian has the same form as described in \pref{eqn:dmft_ham}, however, the $H_0$ term this time represents the tight binding model described by the atomic Wannier basis set, and $H_\text{int}$ reads:
\begin{equation}
\label{eqn:h_int_at}
H_\text{int} = U_\text{at} \sum_{i} \hat{m}_{i}^{\uparrow} \hat{m}_{i}^{\downarrow} 
+ V^\text{intra}_\text{at} \sum_{\langle ij \rangle, \sigma\sigma'} \hat{m}_{i}^{\sigma} \hat{m}_{j}^{\sigma'},
\end{equation}
where now $\hat{m}_i^\sigma$ is the number operator for an atomic orbital on Nb site $i$ and spin channel sigma $\sigma$, and $V^\text{intra}_\text{at}$ is the Nb-Nb interaction inside one trimer. For the DC correction we again employ the fully-localized limit, but this time in the atomic basis:
\begin{equation}
\label{eqn:dc_en_at}
\Sigma^{at}_\text{DC} =  \sum_{i} U  \bigg( M^\text{at}_i -\frac{1}{2} \bigg) \hat{m}^\sigma_{i} \quad ,
\end{equation}
where $M^\text{at}_i = \sum_\sigma \langle m^\sigma_i \rangle $. This terms translates to a downward energetic shift of the occupied atoms and an upward shift of the unoccupied ones.

In our DMFT calculations for the three-atom clusters, we have again omitted the intersite contribution to the double-counting corrections. Given that the material is deep in the Mott regime, the strong Coulomb interaction enforces a local filling of exactly one electron per Nb$_3$ cluster. The subspace of relevant electronic configurations is therefore composed almost exclusively of states with a single occupied atomic orbital per trimer. Within this subspace, the Hamiltonian term for the intersite interaction, $ V^\text{intra}_{at} \sum_{ij}{m_{i}^{\sigma} m_{j}^{\sigma'}}$, always acts as the zero operator and, in fact, we see an extremely weak dependence of our results on the value of $V^\text{intra}_\text{at}$, which makes application of a double counting correction unnecessary.

All our DFT+DMFT calculations are fully charge self-consistent and were performed using \texttt{solid\_dmft}~\cite{Merkel_et_al:2022}, which is part of the \textsc{triqs} library~\cite{Parcollet_et_al:2015}. We solve all the DMFT impurity problems with the continuous-time quantum Monte Carlo solver \textsc{ct-hyb}~\cite{Werner/Millis:2006, Gull_et_al:2011, Seth_et_al:2016} at the inverse electronic temperature close to room temperature of $\beta = (k_B T)^{-1} = 40 \text{eV}^{-1}$. We use 10$^{4}$ warm-up steps and $2\times10^8$ Monte Carlo cycles with 120 steps each. We average over both spin channels to ensure a paramagnetic solution.

From the local Green's function, $G_{\nu \nu'}(\tau)$, where $\nu$, $\nu'$ correspond to different orbitals and $\tau$ is the imaginary time, we obtain the local occupations on a given site (trimer), $n_{\nu \nu'}=G_{\nu \nu'}(\tau=0^-)$, as well as the averaged spectral weight around the Fermi level, $\bar{A}(\omega=0)=-(\beta/\pi)\text{Tr}G(\tau=\beta/2)$. Additionally, we calculate the quasiparticle weight, \mbox{$Z = [1 - \left. \partial \mathrm{Im}\Sigma(i\omega)/\partial(i\omega)\right|_{i\omega \rightarrow 0}]^{-1}$}, from the imaginary part of the local self-energy, $\Sigma$, by fitting a third-order polynomial to the lowest five Matsubara frequencies and interpolating to zero frequency. Finally, we use the maximum-entropy method~\cite{Jarrell/Gubernatis:1996, Kraberger_et_al:2017} to obtain the $k$-averaged spectral functions on the real frequency axis.

\subsection{cRPA calculations}

For the cRPA calculations, norm-conserving pseudopotentials from \textsc{pseudodojo}~\cite{ Hamann:2013, vanSetten_et_al:2018} are used to perform the initial DFT calculations. For these, we use a kinetic energy cutoff of 52\,Ry ($\sim$700\,eV) and 12$\times$52\,Ry for the charge density. We converge the total energies to $10^{-3}$~Ry using a 6$\times$6$\times$3 $k$-point, smearing the occupations with the Marzari-Vanderbilt scheme~\cite{Marzari_et_al:1999} with smearing parameter of 0.001 Ry.

Screened Coulomb interaction parameters are obtained using the cRPA code \textsc{respack}~\cite{Nakamura_et_al:2021}, using the \textsc{wan2respack} interface to \textsc{quantum espresso}~\cite{Kurita_et_al:2023}. We set the polarization function cutoff to 15~Ry and we use a total of 300 bands, setting the upper limit of the highest possible excitations to 25\,eV above the Fermi energy.

\subsection{cRPA results for the \nbx family}
The calculated Coulomb interaction parameters using cRPA~\cite{Aryasetiawan_et_al:2004, Miyake/Aryasetiawan:2008} are given in \pref{tab:crpa_trimer} and \pref{tab:crpa_atom}. We give both the screened and unscreened values for $U$ and $V$ parameters in the $R\bar{3}m$ phase of \nbx compounds in both the trimer and atomic basis sets used in the text. 

In the trimer basis~(\pref{tab:crpa_trimer}), the values of $U_\text{tr}$ and $V_\text{tr}$ change monotonically throughout the series, with increasing $V_\text{tr}/U_\text{tr}$ ratio from $1/4$ to approximately $1/3$. Both the $U$ and the $V$ parameters are strongly screened with increasing ratio of screening across the compounds -- 76\% to 83\% screening for $U$.

Table~\ref{tab:crpa_atom} presents the calculated intratrimer onsite ($U_\text{at}$) and intratrimer ($V^\text{intra}_\text{at}$) and intertrimer ($V^\text{inter}_\text{at}$) intersite interactions in the atomic basis. Similarly to the trimer basis, $U_\text{at}$ and $V_\text{at}$ decrease monotonically across the entire series. Notably, the intertrimer $V^\text{inter}_\text{at}$ interaction remains quite large in the atomic basis, being practically identical to the $V_\text{tr}$ values reported in Table~\ref{tab:crpa_trimer} for the trimer basis.

Our cRPA results for the trimer basis show good agreement with previous works~\cite{Grytsiuk_et_al:2024, Aretz_et_al:2025} performed for the $R\bar{3}m$ and $R3$ structures. As expected, since the underlying structures do not differ except for a slight symmetry-breaking, the results remain comparable. Calculations for the atomic basis have so far been only performed in the monolayer~\cite{Grytsiuk_et_al:2024}, where all the interaction parameters are understandably larger than the ones shown here due to the lack of screening in a 2D structure.

\begin{table}
	\caption{The screened and unscreened interaction parameter values in eV as obtained from cRPA for \nbcl, \nbbr, and \nbi in the trimer basis.}
	\begin{tabular}{p{1.7cm}p{2cm}R{1.3cm}R{1.3cm}}
		\hline
		\hline
		&      (All in eV) & $U_\text{tr}$ & $V_\text{tr}$ \\ \hline
		\multirow{3}{*}{\begin{tabular}[c]{@{}l@{}} \nbcl \end{tabular}} & Screened &  1.46  &  0.38  \\
		& Unscreened   & 6.00 & 2.20 \\
		& Ratio      & 0.24 & 0.17 \\[0.2cm]
		\multirow{3}{*}{\begin{tabular}[c]{@{}l@{}} \nbbr \end{tabular}} & Screened &  1.21 &  0.36  \\
		& Unscreened   &  5.73 &  2.11 \\
		& Ratio      & 0.21 &   0.17 \\[0.2cm]
		\multirow{3}{*}{\begin{tabular}[c]{@{}l@{}} \nbi \end{tabular}} & Screened &  0.85 &  0.28  \\
		& Unscreened   &  5.10 &  1.99 \\
		& Ratio      & 0.17  &  0.14 \\ 
		\hline
		\hline
	\end{tabular}
	\label{tab:crpa_trimer}
\end{table}

\begin{table}
	\caption{The screened and unscreened interaction parameter values in eV as obtained from cRPA for \nbcl, \nbbr, and \nbi in the atomic basis.}
	\begin{tabular}{p{1.3cm}p{2.1cm}R{1cm}R{1cm}R{1cm}}
		\hline
		\hline
		&      (All in eV) & $U_\text{at}$ & $V^\text{intra}_\text{at}$ & $V^\text{inter}_\text{at}$ \\ \hline
		\multirow{3}{*}{\begin{tabular}[c]{@{}l@{}} \nbcl \end{tabular}} & Screened & 2.33  & 1.36 & 0.39 \\
		& Unscreened & 9.70 & 4.96 & 2.02 \\
		& Ratio      & 0.24 & 0.27  & 0.19 \\[0.2cm]
		\multirow{3}{*}{\begin{tabular}[c]{@{}l@{}} \nbbr \end{tabular}} & Screened & 2.00  & 1.10  & 0.35 \\
		& Unscreened & 9.39 & 4.78  & 1.98 \\
		& Ratio     & 0.21  & 0.23 & 0.18 \\[0.2cm]
		\multirow{3}{*}{\begin{tabular}[c]{@{}l@{}} \nbi \end{tabular}} & Screened & 1.52 & 0.78  & 0.27 \\
		& Unscreened   & 8.62  & 4.44 & 1.87 \\
		& Ratio      &  0.18 & 0.18  & 0.14 \\ 
		\hline
		\hline
	\end{tabular}
	\label{tab:crpa_atom}
\end{table}

\begin{figure}
  \includegraphics[width=1\linewidth]{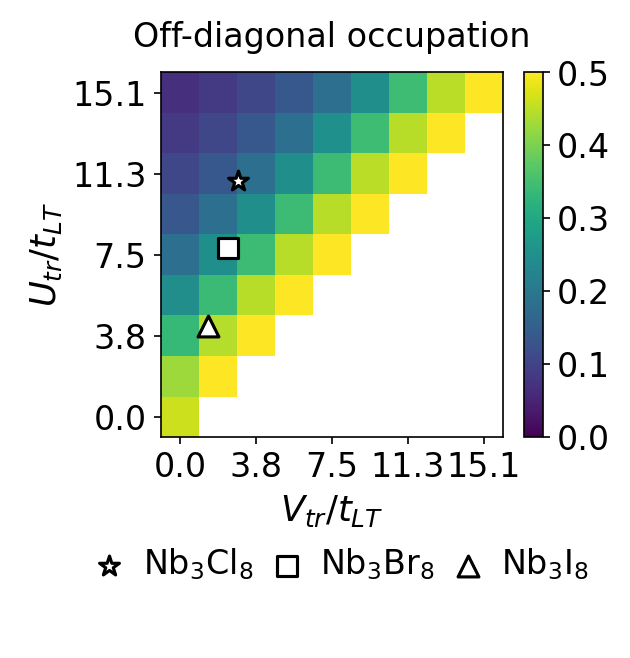}
  \caption{The off-diagonal occupation of the \nbx cluster from DFT+DMFT as a function of $U_\text{tr}/t_\text{LT}$ and $V_\text{tr}/t_\text{LT}$. The cRPA values for \nbcl, \nbbr, and \nbi are shown with a circle, square, and triangle, respectively.}
  \label{fig:dmft-nbx}
\end{figure}

In \pref{fig:dmft-nbx} we show the off-diagonal occupation in analogy to \pref{fig:dmft}(c), indicating the positions of the cRPA-calculated interaction parameters of the \nbx compounds in the trimer basis from \pref{tab:crpa_trimer} (shown as star, square, and triangle, respectively). As we move further down the series, the ratio of $U_\text{tr}/V_\text{tr}$ changes, and we get closer to the singlet insulating regime edge of the diagram, aided by the marked change in $t_{\text{LT}}$ [not shown in \pref{fig:dmft-nbx}]. The increasing proximity to the band insulator is also the reason why the other compounds in the series, \nbbr and \nbi, both feature a band gap already with conventional DFT.

\section{Experimental details}

\textbf{Synthesis of \nbcl.} A mixture of NbCl$_5$ (76.9\,mg, 28.4\,mmol ABCR GmbH, 99.9\%) and Nb (23.1\,mg, 24.9\,mmol, ABCR GmbH, 99.9\%), fused into an evacuated silica tube, was heated with 0.5\,K/min in a Simon-Müller furnace. After heating at 750\,°C for 48 hours, \nbcl was obtained as plate-like crystals (yield: $>95$\,\%).

\textbf{Synthesis of Nb$_3$Br$_4$Cl$_4$.} A mixture of NbBr$_5$ (79.8\,mg, 16.2\,mmol, ABCR GmbH, 99.9\%), NbCl$_5$ (43.8\,mg, 16.2\,mmol ABCR GmbH, 99.9\%) and Nb (26.4\,mg, 28.4\,mmol, ABCR GmbH, 99.9\%), fused into an evacuated silica tube, was heated with 0.5\,K/min in a Simon-Müller furnace. After heating at 750\,°C for 48 hours, Nb$_3$Br$_4$Cl$_4$ was obtained as plate-like crystals (yield: $>90$\,\%).

\textbf{Single-Crystal X-ray Diffraction.} To maintain crystal integrity, separate single crystals were selected for each measurement and gradually cooled to the target temperature at a rate of 2 K/min. Data collection was performed on a Rigaku XtaLAB Synergy-S single-crystal X-ray diffractometer equipped with HyPix-6000HE detector and monochromated Mo-K$_\alpha$ radiation ($\lambda = 71.037$\,pm ) at 100\,K and monocromated Cu-K$_\alpha$ radiation ($\lambda = 154.184$\,pm) at 270\,K. X-ray intensities were corrected for absorption with a numerical method (crystal faces) using CrysAlisPro 1.171.43.121a (Rigaku Oxford Diffraction, 2024). Structures were solved by direct methods (SHELXT) and refined by full-matrix least squares methods performed with SHELXL‐2019/3~\cite{Sheldrick:2015} as implemented in Olex2 1.5~\cite{Dolomanov_et_al:2009}. Detailed crystallographic data can be obtained free of charge via www.ccdc.cam.ac.uk and the CCDC numbers.

\begin{figure}
  \includegraphics[width=0.95\columnwidth]{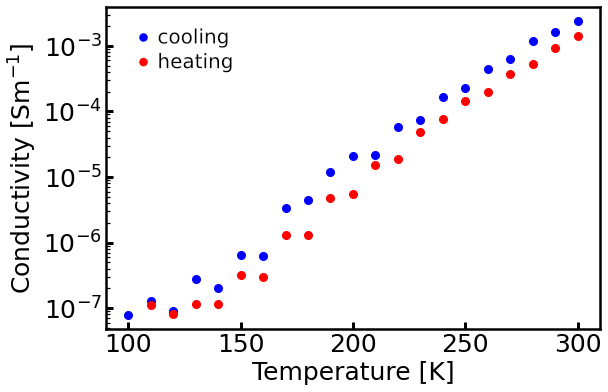}
  \caption{Electrical conductivity of \nbcl versus set temperature in a range of 100\,K to 300\,K. red: cooling; blue: heating}
  \label{fig:elcond}
\end{figure}

\textbf{Electrical Conductivity} Conductivity measurements were performed in a Lake Shore Cryotronics CRX-6.5K probe station with a Keithley 2636B source meter unit. Plate-like crystals of \nbcl were transferred into the chamber under protective gas and contacted with silver paste on a silicon substrate with 770\,nm oxide layer. The conductive silver pads at each end of the crystals were connected to the circuit with gold coated tungsten tips. The chamber was kept under vacuum ($ > 5 \cdot 10^{-5}$ mbar) and the temperature was varied between 20\,K and 300\,K (\pref{fig:elcond} blue: cooling; red: heating). Before each measurement, sufficient time was allowed for the sample to reach the chosen temperature. Two-point conductivity measurements were performed by varying the applied source–drain voltage from -1\,V to 1\,V while detecting the current.

\textbf{DRIFT (Diffuse Reflectance Infrared Fourier Transformation) Spectroscopy.} Samples were measured at room temperature under inert conditions in diffuse reflectance with a Harrick Praying Mantis attachment using a Bruker Vertex 70 infrared spectrophotometer with a deuterated triglycine sulfate (DTGS) detector and KBr beamsplitter. The background spectra were collected using pure dried KI in powder form.

\textbf{Magnetic Studies.} Magnetic susceptibility measurements of crystalline powders and single crystals were performed in gelatine capsules between 5\,K and 300\,K with a Quantum design SQUID magnetometer equipped with a 1802 R/G bridge and a 1822 MDMS controller. 

\subsection{Optical band gap determination from DRIFTS (diffuse reflectance infrared fourier transform spectroscopy)}

The absorption coefficient $F$ was obtained using Kubelka–Munk analysis following:
\begin{equation}
F(R_{\infty}) = \frac{\alpha}{S} = \frac{(1 - R_{\infty})^2}{2R_{\infty}},
\end{equation}
where $R_{\infty} = \frac{R_{\text{sample}}}{R_{\text{standard}}}$ is the reflectance of an infinitely thick specimen, $\alpha$ is the absorption coefficient, and $S$ is the scattering coefficient. For particle sizes greater than the light wavelengths measured, the scattering coefficient is understood to be approximately independent of frequency ($F(R_{\infty}) \sim \alpha$) and therefore $F(R_{\infty})$ could be understood as a "pseudo-absorbance" coefficient~\cite{Wendlandt/Hecht:1966, kubelka:1931, kubelka:1948}.

The band gap determination was performed on the DRIFTS data according to Zanata~\textit{et al.}~\cite{zanatta:2019}. The energy is plotted against the absorption coefficient $\alpha$ and fitted with a sigmoid-Boltzmann function:
\begin{equation}
\alpha(E) = \alpha_{\max} + \frac{\alpha_{\min} - \alpha_{\max}}{1 + \exp\left( \frac{E - E_0^{\text{Boltz}}}{\delta E} \right)},
\end{equation}
where $\alpha_{\min}$ ($\alpha_{\max}$) stands for the minimum (maximum) absorption coefficient; $E_0^{\text{Boltz}}$ is the energy coordinate at which the absorption coefficient is halfway between $\alpha_{\min}$ and $\alpha_{\max}$; and $\delta E$ is associated with the slope of the sigmoid, indicating the energy range over which most optical transitions occur~\cite{zanatta:2019}.

The band gap can then be calculated by the following equation with $n_{\text{dir}}^{\text{Boltz}} = 0.3$ and $n_{\text{indir}}^{\text{Boltz}} = 4.3$:

\begin{equation}
E_g^{\text{Boltz}} = E_0^{\text{Boltz}} - n_{\text{(dir/indir)}}^{\text{Boltz}} \cdot \delta E.
\end{equation}

\begin{figure}
  \includegraphics[width=0.95\columnwidth]{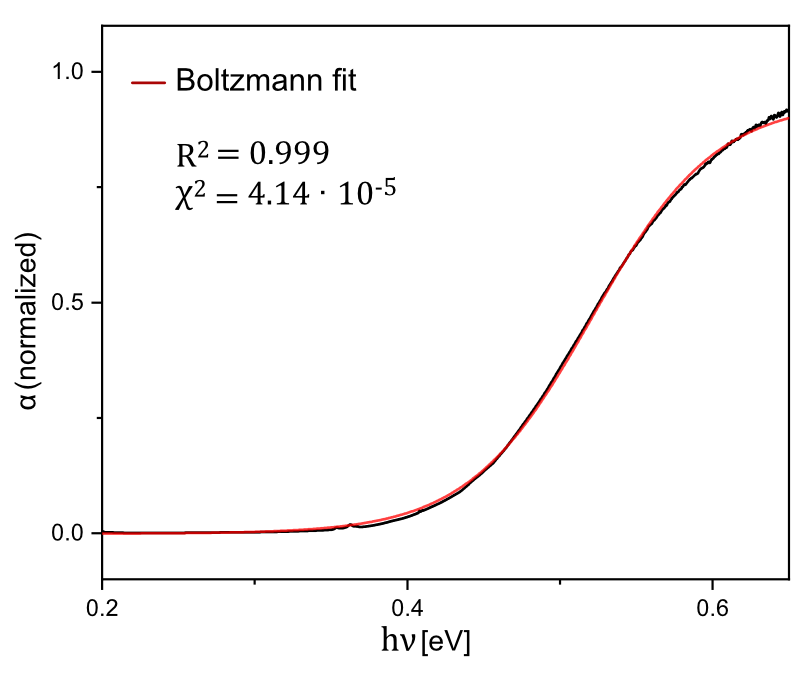}
  \caption{Optical absorption coefficient spectra of \nbcl with the
Boltzmann function used to fit $\alpha$ (normalized).}
  \label{fig:optbandnbcl}
\end{figure}
\begin{figure}
  \includegraphics[width=0.95\columnwidth]{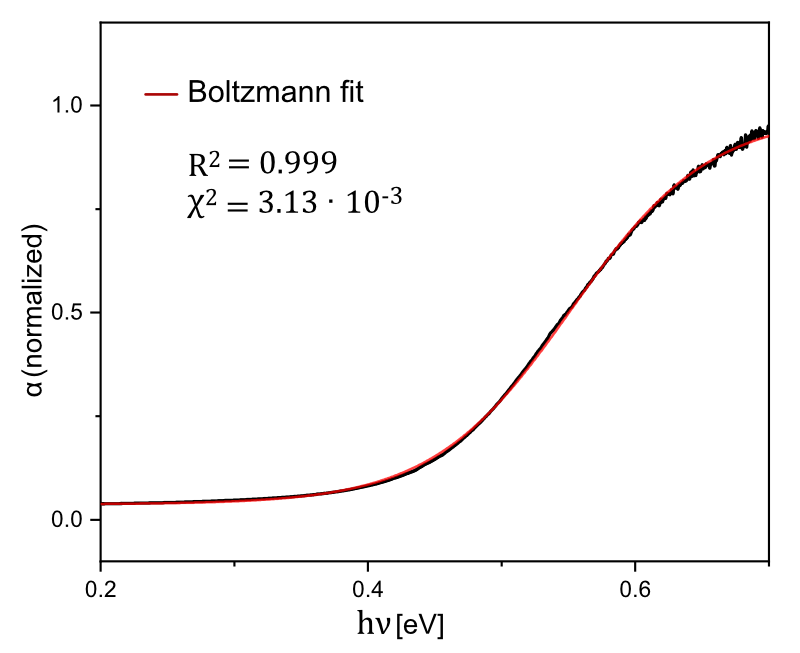}
  \caption{Optical absorption coefficient spectra of Nb$_3$Br$_4$Cl$_4$ with the
Boltzmann function used to fit $\alpha$ (normalized).}
  \label{fig:optbandnbbrcl}
\end{figure}

\subsection{Magnetic measurements}
\begin{figure}
  \includegraphics[width=0.95\columnwidth]{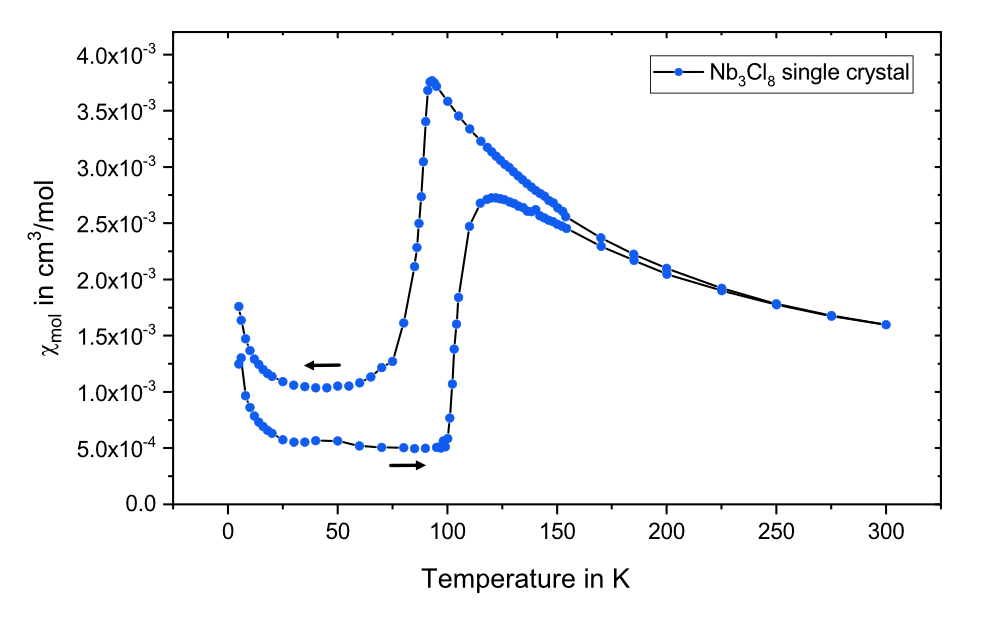}
  \caption{Magnetic susceptibility of single crystals of \nbcl versus set temperature in a range of 5\,K to 300\,K.}
  \label{fig:magSC}
\end{figure}
\begin{figure}
  \includegraphics[width=0.95\columnwidth]{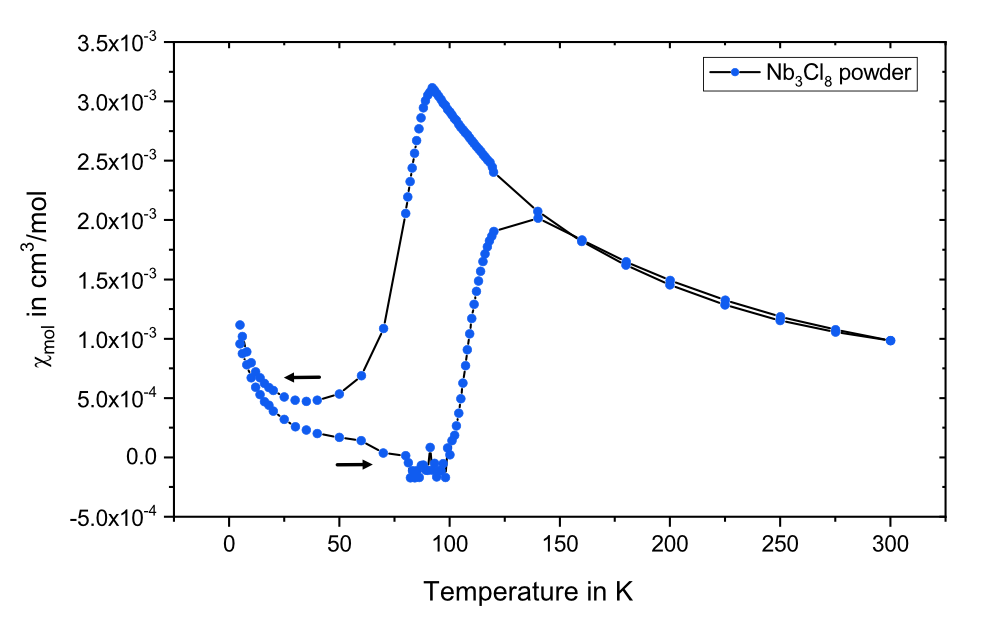}
  \caption{Magnetic susceptibility of a powdered sample of \nbcl versus set temperature in a range of 5\,K to 300\,K.}
  \label{fig:magpowder}
\end{figure}

%%%%%%%%%%%%%%%%%%%%%%%%%%%%%%%%%%%%%%%%%%%%%%%%%%%%%%%%%%%%%%%%%%%%%
%% The appropriate \bibliography command should be placed here.
%% Notice that the class file automatically sets \bibliographystyle
%% and also names the section correctly.
%%%%%%%%%%%%%%%%%%%%%%%%%%%%%%%%%%%%%%%%%%%%%%%%%%%%%%%%%%%%%%%%%%%%%

\clearpage

\bibliography{paperNbCl}

\end{document}